\documentclass[usenatbib]{mn2e}

\pdfoutput=1

\usepackage[english]{babel}
\usepackage[babel]{csquotes}
\usepackage{indentfirst}
\usepackage{setspace}
\usepackage{fancyhdr}
\usepackage{graphicx}
\usepackage{booktabs}
\usepackage{amsmath}
\usepackage{tensor}
\usepackage{placeins}
\usepackage{natbibmnfix}
\usepackage{epsfig}
\usepackage{aas_macros}
\usepackage{color}
 \usepackage{url}

\def\gtsima{$\; \buildrel > \over \sim \;$} 
\def\ltsima{$\; \buildrel < \over \sim \;$} 
\def\gsim{\lower.5ex\hbox{\gtsima}} 
\def\lsim{\lower.5ex\hbox{\ltsima}} 
\def\simgt{\lower.5ex\hbox{\gtsima}} 
\def\simlt{\lower.5ex\hbox{\ltsima}} 
 
\def\Lya{Ly$\alpha$~}

\def\CII{\hbox{C~$\scriptstyle\rm II$}}
\def\OII{\hbox{O~$\scriptstyle\rm II$}}
\def\OIII{\hbox{O~$\scriptstyle\rm III$}}

\pdfoutput=1

\title[\Lya intensity mapping]
  {Observational challenges in \Lya intensity mapping}
\author[P. Comaschi, B. Yue, A. Ferrara]
  {P.~Comaschi$^1$\thanks{Email: paolo.comaschi@sns.it},
  B.~Yue$^1$,
  A.~Ferrara$^{1,2}$\\
  $^1$Scuola Normale Superiore, Piazza dei Cavalieri 7, 1-56126 Pisa, Italy\\
  $^2$Kavli IPMU, The University of Tokyo, 5-1-5 Kashiwanoha, Kashiwa 277-8583, Japan}
\date{Released 2016 Xxxxx XX}

\def\LaTeX{L\kern-.36em\raise.3ex\hbox{a}\kern-.15em
    T\kern-.1667em\lower.7ex\hbox{E}\kern-.125emX}

\begin{document}
	
	\maketitle

\begin{abstract}
Intensity mapping (IM) is sensitive to the cumulative line emission of galaxies. As such it represents a promising technique for statistical studies of galaxies fainter than the limiting magnitude of traditional galaxy surveys. The strong hydrogen \Lya line is the primary target for such an experiment, as its intensity is linked to star formation activity and the physical state of the interstellar (ISM) and intergalactic (IGM)  medium. However, to extract the meaningful information one has to solve the confusion problems caused by interloping lines from foreground galaxies. We discuss  here the challenges for a \Lya IM experiment targeting $z > 4$ sources. We find that the \Lya power spectrum can be in principle easily (marginally) obtained with a 40 cm space telescope in a few days of observing time up to  $z\lsim8$ ($z\sim10$) assuming that the interloping lines (e.g. H$_\alpha$, [\OII],~[\OIII]~lines) can be efficiently removed. We show that interlopers can be removed by using an ancillary photometric galaxy survey with limiting AB mag $\sim 26$ in the NIR bands (Y, J, H, or K). This would enable detection of the \Lya signal from $5<z<9$ faint sources.  However, if a [\CII] IM experiment is feasible, by cross-correlating the \Lya with the [\CII] signal the required depth of the galaxy survey can be decreased to AB mag $\sim 24$. This would bring the detection at reach of future facilities working in close synergy. 
\end{abstract}

\begin{keywords}
 cosmology: observations - intergalactic and interstellar medium - intensity mapping - large-scale structure of universe
\end{keywords}

\section{Introduction}
\label{sec:intro}
One of the key open problems in cosmology is the origin and evolution of galaxies and their stars. In the last decade astonishing technological progresses have allowed  to probe galaxies located within less than one billion year from the Big Bang \citep{2014ApJ...793..115B, 2014ApJ...786..108O, 2015ApJ...804L..30O, 2010ApJ...723..869O, 2008ApJS..176..301O, 2015MNRAS.451..400M}. These searches reveal an early Universe in which complex phenomena were simultaneously taking place, ranging from the formation of supermassive black holes \citep{2012RPPh...75l4901V} to the reionization process,  \citep{2001PhR...349..125B}, along with the metal enrichment by the first stars \citep{2016ASSL..423..163F}. 

High redshift sources are very faint and their detection is remarkably challenging: up to now, less than $1000$ galaxies have been detected at $z \gsim 8$, and among them only a handful are at $z \sim 10$ (e.g. \citealt{2014arXiv1403.4295B,2016arXiv160205199M,2016ApJ...817..120C}). Moreover, it is believed that low-mass galaxies have a dominant role \citep{2011MNRAS.414..847S} in driving reionization, while the most-luminous ones appear to be only rare outliers. Such ultra-faint galaxies are likely to remain undetected even by the next generation observatories, such as JWST\footnote{\url{http://www.jwst.nasa.gov}}, TMT\footnote{\url{http://www.tmt.org}} or E-ELT\footnote{\url{https://www.eso.org/sci/facilities/eelt/}}.

A novel approach has been proposed to overcome the problem and study, at least statistically, the early faint galaxy population. Basically the idea is to trade the ability to resolve individual sources, with a statistical analysis of the cumulative signal produced by the entire population \citep{2005PhR...409..361K, 2016arXiv160203512C}. Intensity mapping (IM, see e.g. \citealt{2010JCAP...11..016V, 2011JCAP...08..010V}) is one implementation of such concept and aims at detecting 3D large scale emission line fluctuations. In the last years this concept has become very popular and several lines have been proposed as candidates. Among these are the HI 21cm \citep{2006PhR...433..181F}, CO \citep{2011ApJ...741...70L, 2008A&A...489..489R, 2014MNRAS.443.3506B} , \CII~ \citep{2012ApJ...745...49G, 2014arXiv1410.4808S, mappingCII}, H$_2$ \citep{2013ApJ...768..130G}, HeII \citep{2015arXiv150103177V} and \Lya \citep{2014ApJ...786..111P, 2013ApJ...763..132S, 2016MNRAS.455..725C} emission lines.

Although IM experiments seem indeed promising, their reliability has not yet been convincingly demonstrated. In particular, continuum foregrounds dominate over line intensity by several orders of magnitude: cleaning algorithms have been developed for 21cm radiation \citep{2006ApJ...650..529W, 2015arXiv150104429C, 2015arXiv150103823W}, but not comparably well understood for other lines \citep{mappingCII}. Moreover, some lines (such as \Lya and FIR emission lines) suffer from line confusion: for example the H$_\alpha$ line ($\lambda_{{\rm H}_\alpha} = 0.6563~\mu$m) if emitted at $z = 0.48$ can be misclassified as a \Lya line emitted at $z=7$ \citep{2014ApJ...785...72G}. We will refer to such intervening sources as \textit{interlopers}.  

Considering that the first generation of instruments devoted to IM are starting to be proposed or funded \citep{2014arXiv1412.4872D, 2016arXiv160205178C, 2014SPIE.9153E..1WC}, it is essential to gain a deeper understanding of the difficulties implied by an IM experiment. This forms the motivation of this work and we will pay particular attention to the \Lya emission line which is the most luminous UV line and one of the most promising candidates for an IM survey in the near infrared (NIR) spectral region.

\Lya emission is associated with UV and ionizing radiation and therefore is strongly correlated with the star formation rate (SFR) in galaxies. Moreover, the reprocessing of UV photons by neutral hydrogen in the IGM also produces \Lya photons.
Some recent works have predicted the power spectrum (PS) of the target line and assessed its observability.
\cite{2014ApJ...786..111P} and \cite{2013ApJ...763..132S} developed analytical models for the \Lya PS and showed that it is at reach of a small space instrument.
\cite{2014ApJ...785...72G} used the model developed by \cite{2013ApJ...763..132S} to study the problem of line confusion, finding that masking bright voxels can represent a viable strategy.
In a similar attempt, \cite{2015MNRAS.452.3408B} pointed out that masking bright voxels is an effective strategy for the removal of the interlopers, but it might jeopardize the recovered line PS, causing loss of astrophysical information. 

A realistic \Lya  model has to deal with all the astrophysics processes (e.g. star formation, radiative transfer) self-consistently. This is rather challenging even for high resolution hydrodynamic simulations. Alternatively, a viable strategy for studying such complex processes is to develop an analytical model that includes all the theoretical uncertainties represented by a few parameters: in this way it is possible to understand easily how the results depends on the unknowns and what is the available parameter space of the problem yielding solution compatible with existing observations. \citet{2016MNRAS.455..725C} (hereafter CF16) developed an analytical model for diffuse \Lya intensity and its PS, with a focus on IM at the epoch of reionization (EoR). The model is observation-driven and it includes the most recent determinations both for galaxies and IGM. They associated dust-corrected UV luminosity to dark matter halos by the abundance matching technique \citep{2009ApJ...696..620C, 2010ApJ...717..379B, 2004MNRAS.353..189V}, using the LF from the Hubble legacy fields \citep{2015ApJ...803...34B}, and the UV luminosity spectral slope in \citet{2014ApJ...793..115B}. Then using a template spectral energy distribution (SED) from \texttt{starburst99}\footnote{http://www.stsci.edu/science/starburst99/docs/default.htm} \citep{1999ApJS..123....3L, 2005ApJ...621..695V, 2014ApJS..212...14L} and the Calzetti extinction law \citep{2000ApJ...533..682C} they were able to model self-consistently the interaction of ionizing photons with the  interstellar medium (ISM) and the IGM, calibrating the poorly constrained parameters in order to have a realistic reionization history \citep{2015arXiv150201589P, 2006AJ....132..117F}.
 
CF16 found that for \Lya absolute intensity is dominated by recombinations in ISM, and Lyman continuum absorption and relaxation in the IGM, with the latter being about a factor 2 stronger. However, intensity fluctuations are mostly contributed by the ISM emission on all scales $<100~h^{-1}$Mpc. Such scale essentially corresponds to the distance at which UV photons emitted by galaxies are redshifted into \Lya resonance.

We present in the following a feasibility study of a \Lya IM survey based on CF16 results. In particular, we tackle the problem of (i)  required sensitivity; (ii) suppression of line confusion through interlopers removal; (iii) detectability of the cross-correlation with the \CII~line. The paper is organized as follows: in Sec. \ref{sec:recap} we compute in a general way the signal-to-noise ratio (S/N) of an IM observation; in Sec. \ref{sec:fore} we model the sensitivity of an intensity mapper and compute the S/N of an observation; in Sec. \ref{sec:inter} we analyse the problem of line confusion. Sec. \ref{sec:cps} contains a study of the cross-correlation between \Lya and \CII~ emission and of the S/N of a realistic observation\footnote{We assume a flat $\Lambda$CDM cosmology compatible with the latest Planck results: $h = 0.677$, $\Omega_m = 0.31$, $\Omega_b = 0.049$, $\Omega_\Lambda = 1 - \Omega_m$,  $n = 0.97$, $\sigma_8 = 0.82$ \citep{2015arXiv150201589P}.}.

\section{Signal Power spectrum}
\label{sec:recap}
In this Section we derive the PS (auto-correlation PS and cross-correlation PS) of the measured intensity fluctuations and its variance, with an approach similar to \cite{2010JCAP...11..016V}. For simplicity we assume that the detected intensity includes three components:  (i) the signal; (ii) the instrumental white noise; (iii) the interloping lines which are redshifted to the same frequency as the signal line, namely 
\begin{equation}
\label{itot}
  I(\Omega, \nu) = I_\alpha(\Omega, \nu) + I_{\rm N} + \sum_i I^i_f(\Omega, \nu).
\end{equation}
Throughout work we will neglect the possible presence of continuum foregrounds, assuming that they can be easily removed thanks to the smoothness of the frequency spectrum \citep{2006ApJ...650..529W, 2015arXiv150104429C, 2015arXiv150103823W}.

Note that comoving coordinates are related to angle and frequency displacement from an arbitrary origin, $\mathbf{x}^0$, as follows:
\begin{gather}
x_1, x_2 = \chi(z_\alpha) \Delta \theta + x^0_{1},x^0_2\\
x_3 = \frac{d\chi}{d\nu}\Delta \nu + x^0_{3}
\end{gather}
where $\chi(z_\alpha)$ is the comoving distance from the observer to the signal, $d\chi/d\nu = c(1+z_\alpha)[H(z_\alpha)\nu]^{-1}$, $(\Delta \theta, \Delta \nu)$ are the displacements in angle and frequency from the origin ${\bf x}^0$ (center of the survey). In this process a subtlety arises \citep{2010JCAP...11..016V, 2014ApJ...785...72G}  because $I^i_f$ is not emitted at $z_\alpha$. Therefore, in that term we should consider coordinates that are the projection at $z_\alpha$ of the real coordinates at $z_i$:
\begin{multline}
I({\bf x}) = I_\alpha({\bf x}, z_\alpha) + I_{\rm N} + \\
+ \sum_i I^i_f\left(x_1\frac{\chi(z_i)}{\chi(z_\alpha)}, x_2\frac{\chi(z_i)}{\chi(z_\alpha)}, x_3 \frac{(1+z_i)H(z_\alpha)}{(1+z_\alpha)H(z_i)}, z_i\right)
\end{multline}
where $1+ z_i = (1+z_\alpha)\lambda_\alpha/\lambda_i$. 

When considering the Fourier transform of the fluctuations, this projection introduces (i) a global extra factor that multiplies the PS; (ii) anisotropies due to the different projection of modes along and across the line of sight; (iii)  a loss of correspondence between comoving and observed $k$-modes: 
\begin{equation}
\label{deltaitot}
 \delta I({\bf k}) = \overline I_\alpha \langle b \rangle_\alpha \delta^\alpha_{\bf k} + \delta^{N}_{\bf k} + \sum_i C(z_i) \overline I_f^i \langle b \rangle_i \delta^{i}_{{\bf k'}({\bf k})};
\end{equation}
where $\overline I$ and $\langle b \rangle$ with each subscript are the mean intensity and halo luminosity weighted mean bias of each line; $\delta^{N}_{\bf k}$ is the instrumental noise (see Sec. \ref{sec:fore}). The global extra factor is
\begin{gather}
\label{interamp}
C(z_i) = \left( \frac{\chi(z_\alpha)}{\chi(z_i)} \right)^2 \frac{(1+z_\alpha)H(z_i)}{(1+z_i)H(z_\alpha)}; \\
{\bf k'}({\bf k}) = \left( k_1 \frac{\chi(z_\alpha)}{\chi(z_i)},  k_2 \frac{\chi(z_\alpha)}{\chi(z_i)}, k_3 \frac{(1+z_\alpha)H(z_i)}{(1+z_i)H(z_\alpha)}  \right).
\end{gather}

From the above equations, the PS of the measured intensity fluctuations becomes
\begin{equation}
\label{psline}
P({\bf k}) = \langle \delta I({\bf k}) \delta I^*({\bf k}) \rangle = P_\alpha({\bf k}) + P_{\rm N} + \sum_i P_f^i 
\end{equation}
%
where 
\begin{gather}
 P_\alpha({\bf k}) = {\overline I_\alpha }^2\langle b \rangle_\alpha^2 P_{\rm dm}({\bf k},z_\alpha),\nonumber \\
 P_{\rm N} = \langle \delta^{\rm N} \delta^{\rm N} \rangle,\nonumber \\
 P_f^i({\bf k}) = C(z_i) ({\overline I^i_f})^2 \langle b \rangle^2_i P_{\rm dm}({\bf k'},z_i)\nonumber,
\end{gather}
in deriving the last line the relation $\langle \delta_{{\bf k'}({\bf k})} \delta_{{\bf p'}({\bf p})} \rangle = C^{-1} P_{\rm dm}({\bf k'}) \delta^3({\bf k} - {\bf p})$ is used and $P_{\rm dm}$ is the dark matter PS.
The noise component $P_{\rm N}$ is well known and easily subtracted; the interlopers power spectrum, $P_f^i({\bf k})$, is however unknown and yet must be removed in order to extract the astrophysical PS signal. 

The variance of $P(\bf k)$ is
\begin{equation}
\label{ptotvar}
\sigma^2_{P}({\bf k}) =\delta P^2({\bf k})= \langle \left( \delta I({\bf k}) \delta I^*({\bf k}) \right)^2 \rangle - \langle \delta I({\bf k}) \delta I^*({\bf k}) \rangle^2.
\end{equation}
Using that fact that noise and interloping lines only correlate with themselves, and that $\langle \mid \delta_{\bf k}\mid\rangle^4 = 2 \sigma_{\bf k}^4$ and $\langle \vert \delta^{\rm N} \vert^4 \rangle = 2 P^2_{\rm N}$, it is easy to prove (see Appendix \ref{app:fullcalc} for the full calculation)
\begin{equation}
\label{varianceline}
\sigma^2_{P}({\bf k})  = \left[{P_{\alpha}({\bf k})} + P_{\rm N} + \sum_i P_f^i({\bf k})\right]^2.
\end{equation}
From this equation we can see that the variance depends strongly on the detector noise and on the PS of the interloping lines. 

In case the PS is isotropic, $P({\bf k}) = P(k)$, several independent modes can be combined to reduce the PS variance at given $k$:
\begin{equation}
\label{psisosum}
 P(k) = \left(\sum_{\bf k} \frac{1}{\sigma^2_{P}({\bf k})} \right)^{-1} \sum_{\bf k} \frac{P({\bf k})}{\sigma^2_{P}({\bf k})},
\end{equation}
where the sum is over all the modes with $\vert{\bf k}\vert  = k$.

In order to estimate the S/N we have to consider the PS variance due to the finite survey volume and resolution. In this case the probed $k$-modes are discrete and multiples of $(2\pi  / L_1, 2\pi /L_2, 2\pi /L_3)$, where $L_1, L_2, L_3$ are the dimensions of the survey volume. Suppose the survey has a resolution $l_\parallel$ and $l_\perp$ along and perpendicular to the line-of-sight (generally $l_\parallel \gg l_\perp$), respectively. Then only modes satisfying $2\pi/L_1,2\pi/L_2<k_1, k_2 < 2\pi / l_\perp$ and $2\pi/L_3<k_3 < 2 \pi/ l_\parallel$ are accounted.

Sometimes it is useful to estimate the total PS variance and S/N for all modes with $k_{\rm min} <k<k_{\rm max}$ \citep{2014ApJ...786..111P}: 
\begin{gather}
\langle \sigma_{P}^2 \rangle = \left( \int \frac{d^3k}{\Delta k^3} \frac{1}{\sigma^2_{P}({\bf k})}\right)^{-1}; \\ 
\langle \label{sn}(S/N)^2 \rangle= \int \frac{d^3k}{\Delta k^3} \left(\frac{P({\bf k})}{\sigma_{P}({\bf k})} \right)^2,
\end{gather}
where $\Delta k^3 = (2\pi)^3/V_{\rm s}$ is the $k$-space volume occupied by each discrete mode and the integral is over all 
wavenumbers with $ k_{\rm min} < \mid {\bf k} \mid < k_{\rm max}$, $k_1, k_2 < 2\pi / l_\perp$ and $k_3 < 2 \pi/ l_\parallel$. 

The contamination in the auto-correlation PS (Eq. \eqref{psline}) could be suppressed by cross-correlating different measurements targeting two different signals, $\alpha$ and $\beta$, that are contaminated by uncorrelated interloping lines
\citep{2010JCAP...11..016V}. The cross-correlation PS is \citep{2010JCAP...11..016V}:
\begin{equation}
  P_{\alpha,\beta}({\bf k}) = \langle I_\alpha \rangle \langle b \rangle_\alpha \langle I_\beta \rangle \langle b \rangle_\beta P_{\rm dm}({\bf k}),
\end{equation}
where only the signal term is left as noise and interloping terms are uncorrelated for $\alpha$ and $\beta$. Nevertheless, noise and interloping lines increase the variance:
\begin{multline}
  \label{crvar}
  \sigma^2_{P\alpha\beta} = \frac{1}{2}\left[ P^2_{\alpha\beta} + \right.\\
+ \left. \left(P_\alpha + P_{N,1} + \sum_i {P^i_{f,1}}  \right)\left( P_\beta + P_{N,2} + \sum_i {P^i_{f,2}}\right)   \right],
\end{multline}
where the subscripts $1, 2$ represent the qualities in the two measurements respectively. We will apply this suppression method to our model and discuss more specific details in Sec. \ref{sec:cps}.

\section{Line Detectability}
\label{sec:fore}

We start by assessing first the detectability of the \Lya PS without considering the interlopers contamination. Our discussions are based on different setup parameters of a small space telescope that can map efficiently a large sky area in the visible (corresponding to $2.2 < z_\alpha < 4.8$) and NIR ($z>4.8$) spectral bands. We do not aim at proposing a optimal setup of such instrument, but rather at understanding to what extent the \Lya IM is a viable tool for studying high-$z$ galaxies.

The size of the voxel is one of the most relevant factors for detectability. The voxel size along the line-of-sight is given by $l_\parallel  =\frac{dl}{dz}\Delta z = \frac{c (1+z)}{H(z) R} $; in the perpendicular direction it is instead  $l_\perp = \chi(z) \theta_\mathrm{min}$. As such, it depends on the spectral resolution, $R$, and angular resolution, $\theta_{\rm min}$, of the telescope. The choice of an optimal $l_\parallel$ and $l_\perp$ is crucial: a small voxel results in a smaller volume loss following interlopers removal but requires a longer time to complete a survey for a given area; large voxels suffer from the opposite problem.  Moreover, as we will discuss in Sec. \ref{sec:inter}, there are additional limitations imposed by the ancillary imaging-survey used to identify the interlopers. The latter sets the minimum voxel size to the precision of the redshift measurement (i.e. typically $\approx 0.05(1+z)$ for photometric surveys) along the line-of-sight. It is necessary to find a balanced choice that is specific to the IM experiment configuration and goals.
 
Our fiducial instrument has a $\theta_\mathrm{min} = 6 ~$arcsec beam FWHM (full width at half maximum), a spectrometer with resolution $R = \lambda/\Delta \lambda = 100$ and a survey area of $ 250~\mathrm{deg}^2$ \citep{2014ApJ...786..111P, 2013ApJ...763..132S, 2014arXiv1412.4872D, 2016arXiv160205178C}. Therefore the sample space has voxels with 
$\Delta l_\parallel=$ 35.3 and 28.1 Mpc, and $\Delta l_\perp =$ 214 and 257~kpc, for $z=4$ and 7 respectively. 

In our setup, the voxel size is always larger than the galaxy correlation length (typically $\approx 1$~Mpc$^3$); therefore we expect that each voxel contains several galaxies. Also, as $\Delta l_\parallel \gg \Delta l_\perp$ (typically $\approx 20-30$~Mpc vs. $\approx 200-300$~kpc), only transversal modes contribute to the PS measurement at  $k > 0.1~h {\rm Mpc}^{-1}$.

Another relevant crucial point is the instrumental noise. A space telescope is usually background limited, i.e. the noise level\footnote{We ignore dark current and readout noise as they depend strongly on survey implementation; This approximation is safe at least for instruments similar to SPHEREx (M. Zemcov, private communication)} is set by Poisson fluctuations of the background light. For for \Lya observations the most important background is the Zodiacal Light (ZL). In this case,
\begin{multline}
\label{sigmaalpha}
\sigma_N \approx  1.37 \, \epsilon^{-1/2} \,  {\rm nW m}^{-2}{\rm sr}^{-1} \times   \\
 \left[\frac{\mu {\rm m}}{\lambda}\frac{R}{100}\frac{ I_{\rm ZL}  }{10^3  {\rm nW m}^{-2}{\rm sr}^{-1}} \frac{0.126 {\rm m}^2}{\pi D^2}  
 \frac{8.5\times 10^{-10}{\rm sr}}{\Omega_{\rm pix}} \frac{10^5{\rm s}}{t_{obs}}\right]^{1/2},
\end{multline}
where $I_{\rm ZL}=\nu I_\nu \approx 10^{2-3}  {\rm nW m}^{-2}{\rm sr}^{-1} $  is the typical ZL flux in the relevant frequency range \citep{2005PhR...409..361K, 2014arXiv1412.4872D}. An efficiency $\epsilon$ is added to account for the photons loss by mirrors and integral field unit (IFU),  here assumed conservatively to be $\epsilon=0.25$. 

\subsection{Power spectrum observations}
\label{ssec:foreobs}
The first generation of intensity mappers are likely to have a limited  S/N that will allow only to probe EoR \Lya fluctuations power-spectrum. In this Section we discuss an instrument designed for this aim. Nevertheless in the future more powerful instruments could undertake tomographic observations and we will discuss such possibility in Sec. \ref{ssec:tomo}.
	
The PS of the instrumental noise is  
\begin{equation}
\label{pnoise}
P_N(z) = \sigma_N^2 V_\mathrm{pix};
\end{equation}
 where $\sigma_N$ is from Eq. \eqref{sigmaalpha} and $V_{\rm pix}$ is the comoving voxel volume. We then compute the S/N using $\sigma^2_P =(P_\alpha+P_N)^2$ and Eq. \eqref{sn} (we use $\Delta z = 1$ for the survey volume $V_s$ and divide $k$-space in $k$-bins with $\Delta k= 1.2 k$).

A telescope with a FOV of $\approx 20-30 ~{\rm deg}^2$ (similar to the proposed SPHEREx, see \citealt{2014arXiv1412.4872D}) can observe a field of $200-250 ~{\rm deg}^2$ in two years, with exposure time $t_{\rm obs} \approx 10^7$ s per pointing. Here we consider a more conservative setup with $t_{\rm obs} = 10^5$~s, it is more feasible as it only takes several days to complete the survey.

\begin{figure}
\vspace{+0\baselineskip}
{
\includegraphics[width=0.45\textwidth]{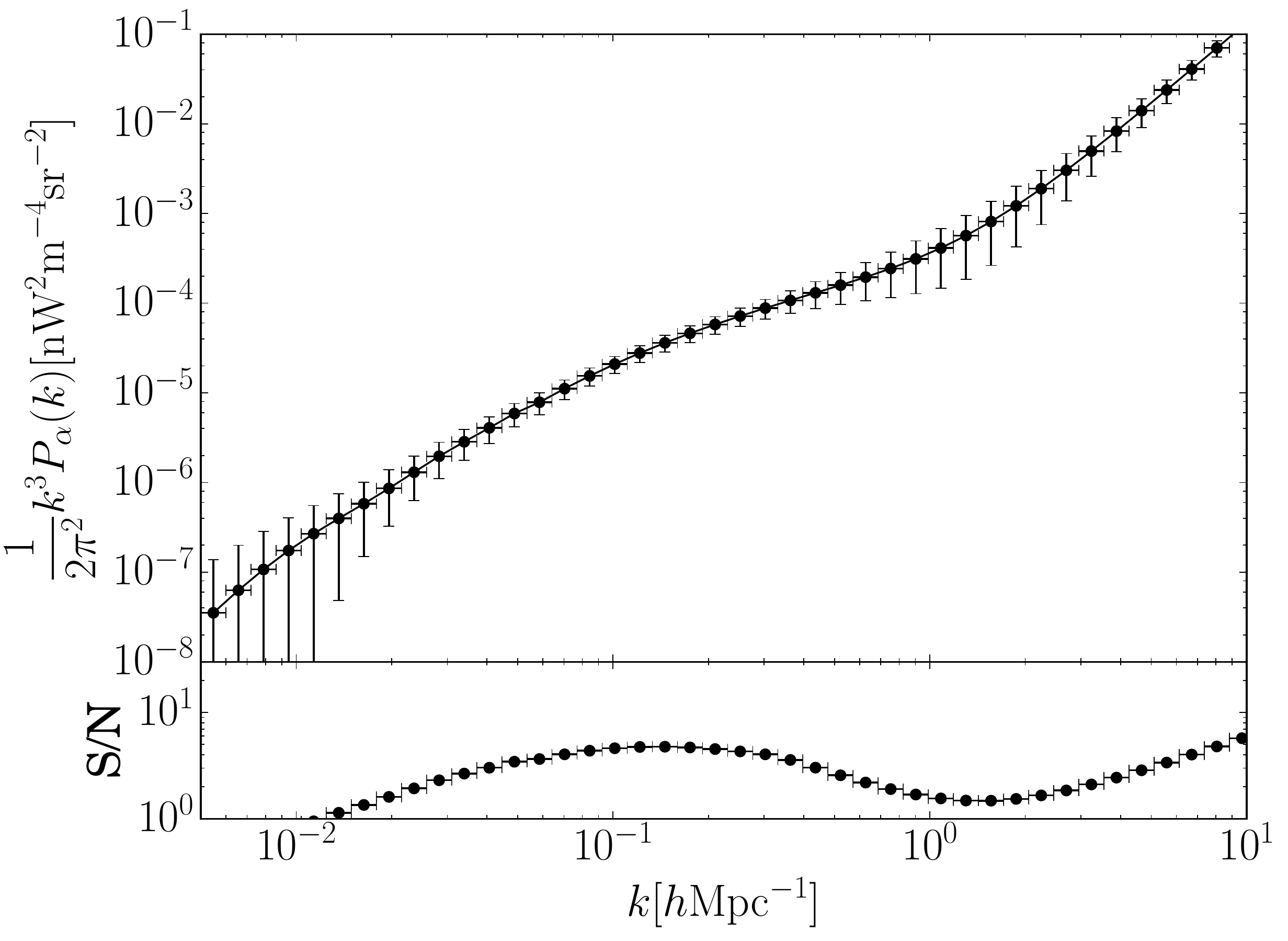}
}
\caption{\textbf{Top:} Predicted \Lya power-spectrum from $z=7$ with errorbars; \textbf{Bottom:} S/N in each $k$-space bin. The S/N is computed for a background-limited NIR telescope with diameter $D=0.4$~m, angular and spectral resolution $(\delta \theta = 6~{\rm arcsec}, R=100)$, exposure time $10^5$~s per pointing and a survey area of $250~{\rm deg}^2$. Each bin has a width $\Delta k = 1.2 k$. 
}
\vspace{-1\baselineskip}
\label{fig:ps_err}
\end{figure}
 
Fig. \ref{fig:ps_err} shows the \Lya PS from $z=7$, and the corresponding S/N assuming $10^5$~s exposure time per pointing and a $250~{\rm deg}^2$ survey area\footnote{This survey set-up is rather conservative; a deeper survey should be possible.}. The S/N is proportional to the number of probed modes: it scales as $k^3$ for bins with $k \lsim 0.1~h{\rm Mpc}^{-1}$ and as $k^2$ for smaller scales, due to the limited spectroscopic resolution. This transition generates a decreasing S/N for $ 0.1~h{\rm Mpc}^{-1} \lsim k \lsim 1~h{\rm Mpc}^{-1}$, where the PS is steeper than $k^{-2}$. Above $k \sim 1~h{\rm Mpc}^{-1}$  the S/N increases again because shot noise dominates and PS is constant. However, as discussed in CF16, shot noise on the $\sim$Mpc scale might be suppressed by \Lya diffusion in the IGM, and therefore the S/N can be overestimated in that range. We conclude that \Lya intensity mapping is best suited to study fluctuations in the linear regime on scales $\sim 10$~Mpc. These results are encouraging because they show that \Lya IM from the late EoR can be detected, provided that continuum foregrounds and low redshift interlopers can  be efficiently removed.

\begin{figure}
\vspace{+0\baselineskip}
{
\includegraphics[width=0.45\textwidth]{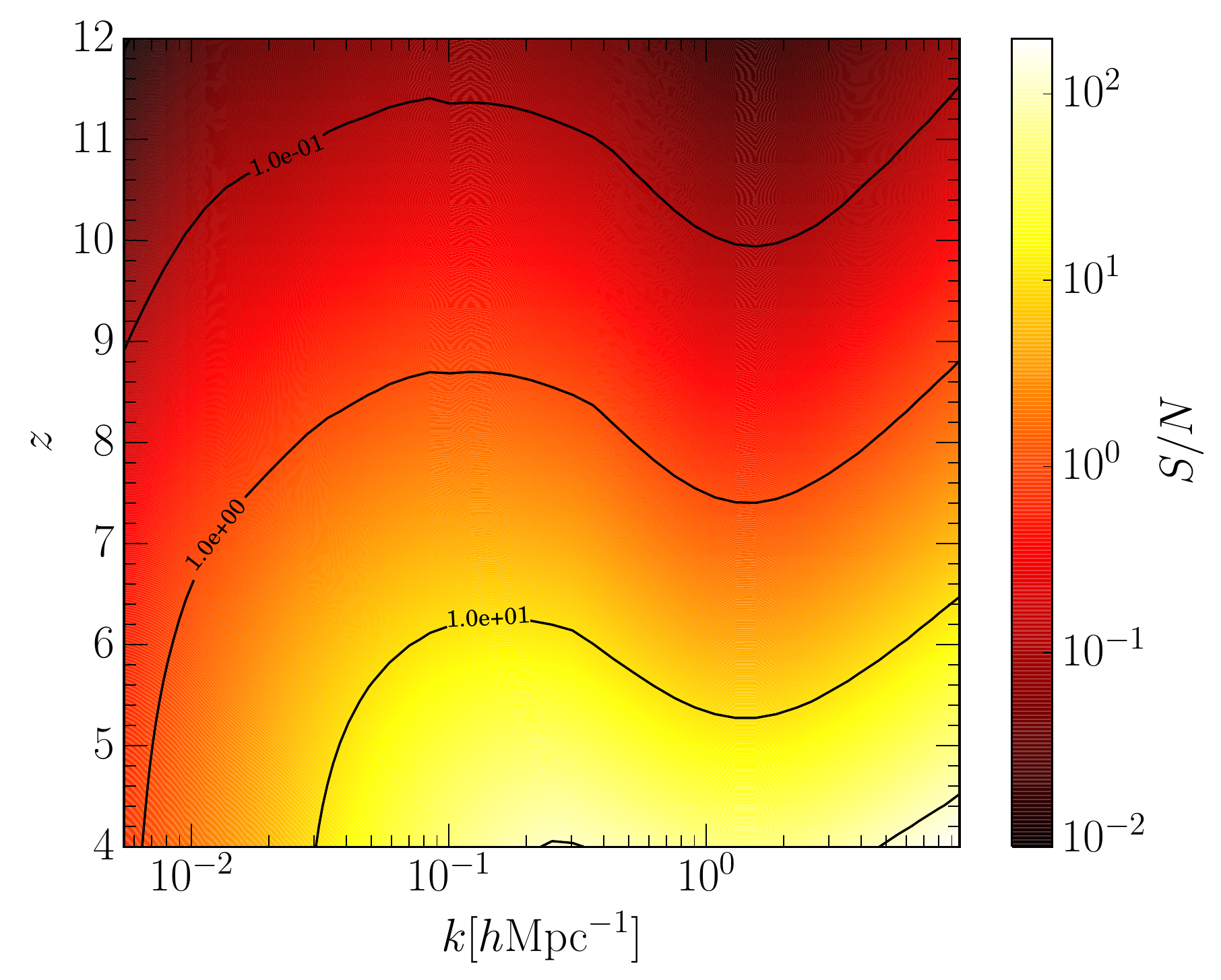}
}
\caption{The S/N of the observed PS as function of redshift $z$ and wave-number $k$; the observational setup is the same of Fig. \ref{fig:ps_err}.}
\vspace{-1\baselineskip}
\label{fig:sn_contour_I}
\end{figure}

Fig. \ref{fig:sn_contour_I} shows a more general dependence of S/N on wavenumber $k$ for \Lya signals coming from different redshifts. The observational setup is the same as in Fig. \ref{fig:ps_err}. We find that the \Lya PS is accessible to this kind of observations at least for the late EoR (i.e. S/N $\approx 5$  at $k = 0.1$~$h$Mpc$^{-1}$  for $z \sim 7$). 

\begin{figure}
\vspace{+0\baselineskip}
{
\includegraphics[width=0.45\textwidth]{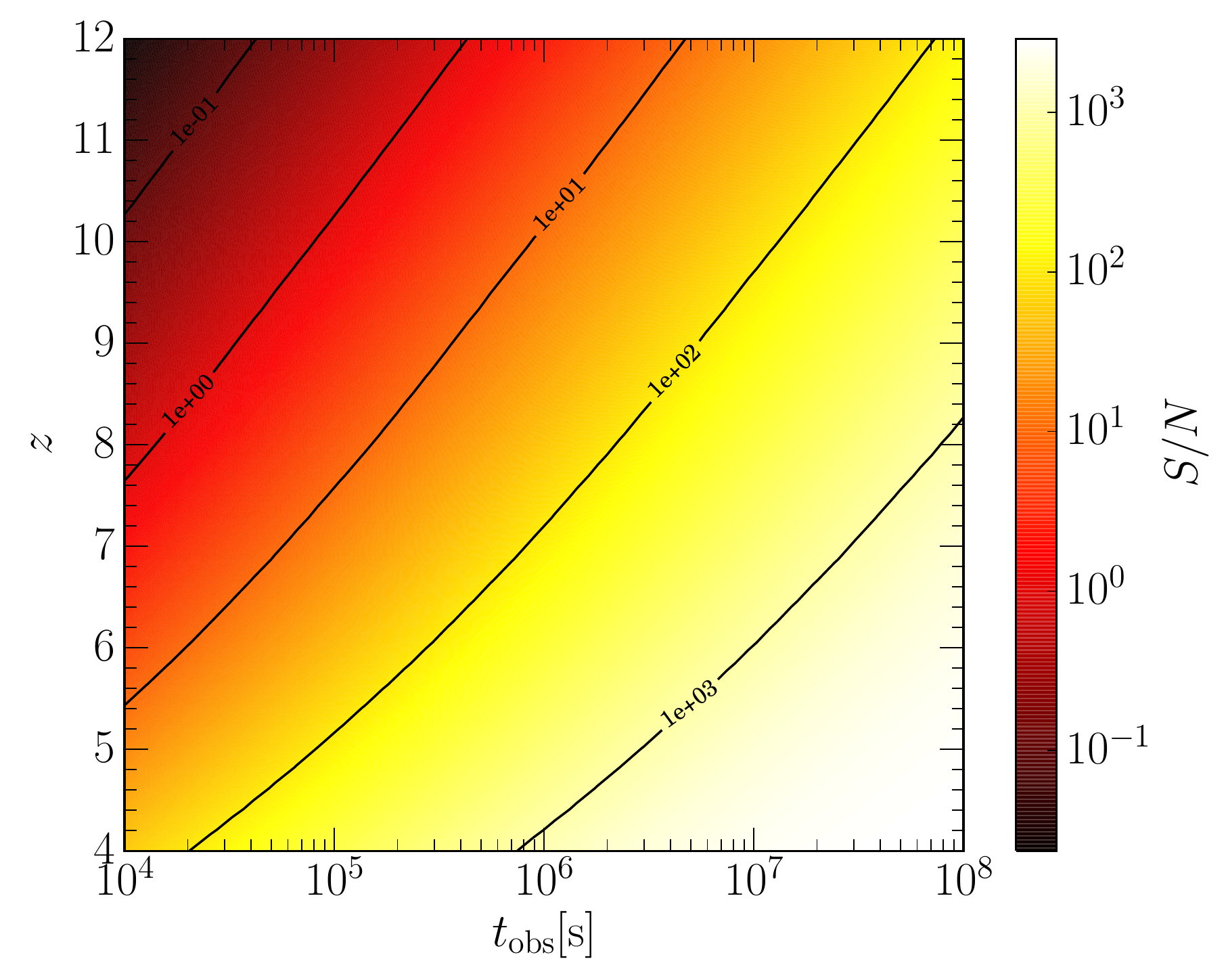}
}
\caption{Total S/N for detection of \Lya fluctuations; the observational setup is the same of Fig. \ref{fig:ps_err}.}
\vspace{-1\baselineskip}
\label{fig:sn_contour_II}
\end{figure}

We then investigate how the detectability depends on varying exposure time $t_{\rm obs}$, using the total S/N, computed using Eq. \eqref{sn}, in the range $ 5\times10^{-3}~h{\rm Mpc}^{-1} < k < 2~h{\rm Mpc}^{-1}$ as the indicator. The results are plotted in Fig. \ref{fig:sn_contour_II}. From there we see that a detection of \Lya PS with low S/N is at reach even at $z > 7$. 
This formalism also allows us to find the best observational strategy for a PS observation: given a fixed total observing time we want to find the optimal exposure time per pointing. Considering only the instrumental noise, from Eq. \eqref{sn}, \eqref{sigmaalpha} and \eqref{pnoise}  we have
\begin{equation}
 {\rm (S/N)}^2 \propto \frac{A_{\rm surv}}{\sigma_N^2}.
\end{equation}
Since $A_{\rm surv} \propto t_{\rm obs}^{-1}$ and $\sigma_N^2 \propto t_{\rm obs}^{-1}$, the S/N does not depend on the depth of the survey as long as the cosmic variance term negligibly appears in Eq. \eqref{varianceline}. In other words the best strategy for an IM experiment is to carry out a shallow, however large area survey.

In practice, though, the optimal $t_{\rm obs}$ is set by the technical implementation of the survey, which should take into account the following limitations: (i) $t_{\rm obs}$ cannot be shorter than, or even comparable to, the instrumental pointing time; (ii) with a large survey area it is impossible to avoid sky regions with higher foregrounds; (iii) as we will discuss in  Sec. \ref{sec:inter}, the IM survey might need deep ancillary galaxy surveys for interloper removal, and therefore the data available for final analysis is limited to the overlapping sky regions.

\subsection{Tomography}
\label{ssec:tomo}

Alternatively, an IM experiment allows us to make tomographic maps of the \Lya intensity, although only the low-$z$ part of the signal is accessible to fiducial space telescope design introduced above.

In CF16 we found that the mean \Lya intensity at $z=4$ is $I_\alpha\approx 0.1~{\rm nW m}^{-2}{\rm sr}^{-1}$. At the same redshift the dark matter field has a fluctuations level  of $\sigma_{\rm dm} \approx 0.23$ on 10 Mpc scales, and the mean \Lya bias $\langle b \rangle_\alpha \approx 3$. Therefore if the survey has voxels of volume $(10~{\rm Mpc})^3$, corresponding to $R \approx 350$ and $\Delta \theta_{\rm pix} = 4.7$arcmin at $z=4$, the $1\sigma$ \Lya fluctuations level is
\begin{equation}
\sigma_\alpha =  I_\alpha  \langle b \rangle_\alpha  \sigma_{\rm dm}   \approx 0.07~ {\rm nW}/{\rm m}^2/{\rm sr},
\end{equation}
which is larger than the noise level  $\sigma_N\approx 0.04$~${\rm nW m}^{-2}{\rm sr}^{-1}$ in Eq. \eqref{sigmaalpha} for $t_{\rm obs} = 10^6$~s. Therefore even this small intensity mapper can observe directly the spatial fluctuations of \Lya emission from low redshift galaxies, although with a modest S/N.

\begin{figure}
\vspace{+0\baselineskip}
{
\includegraphics[width=0.45\textwidth]{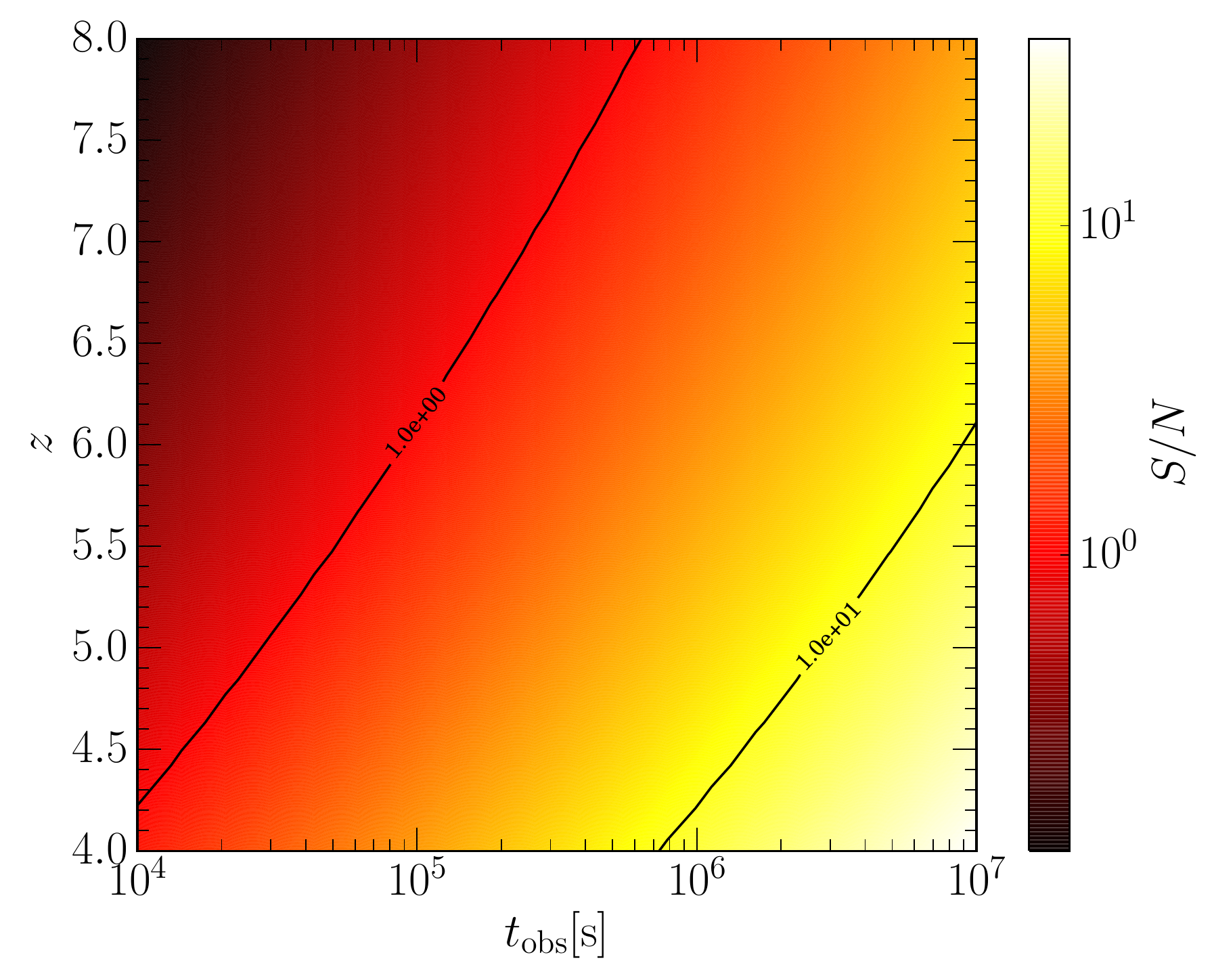}
}
\caption{The S/N of tomographic observations for a voxel size (10 Mpc)$^3$. The S/N is computed for a background-limited NIR telescope with diameter $D=2$~m, angular and spectral resolution $(\delta \theta = 4.7~{\rm arcsec}, R=350)$.}
\vspace{-1\baselineskip}
\label{fig:tomo}
\end{figure}

The tomographic observation of the \Lya signal from the EoR is more challenging, as the \Lya intensity drops by one order of magnitude.  Thus a tomographic map of the EoR signal requires  a more powerful instrument.  
Fig. \ref{fig:tomo} shows the ${\rm S/N}=\sigma_\alpha/\sigma_{\rm N}$ as a function of $z$ and $t_{\rm obs}$ for a 2 m space telescope and same voxels of $(10~{\rm Mpc})^3$ volume. The observation requires an integration time of at least few months and even so it will be only feasible for the late stages of the EoR. The experiment can be even more challenging once the confusion by interloping lines such as H$_\alpha$ and [\OII]~ that dominate over \Lya emission are accounted for. 

\section{Interloping lines}
\label{sec:inter}
Low redshift emission lines could significantly  contribute to the observed intensity fluctuations (see Eq. \eqref{psline}). Particularly important for \Lya experiments are the H$\alpha$ ($0.6563~\mu m$), [\OIII]~ ($0.5007~\mu m$) and [\OII]~ ($0.3727~\mu m$) \citep{2014ApJ...785...72G, 2014ApJ...786..111P} lines. Their power spectra may dominate the \Lya signal, and are distorted and amplified due to coordinate projection effects. Their contribution must therefore be accurately removed from the received flux. In what follows we investigate the power spectra of these interloping lines and suggest a technique to remove them.

\subsection{Power spectra of interloping lines}\label{ssec:interps}

\begin{table}
\centering
\begin{tabular}{ccccc}
\hline
H$_\alpha$  & $z$ & $\phi$ & $L_\star$ & $\alpha$ \\
L07 & 0.24 & $-2.98 \pm 0.40$ & $41.25 \pm 0.34$ & $-1.70 \pm 0.10$\\
    & 0.4  & $-2.40 \pm 0.14$ & $41.29 \pm 0.13$ & $-1.28 \pm 0.07$\\
D13 & 0.25 & $-2.43 \pm 0.19$ & $40.83 \pm 0.18$ & $-1.03 \pm 0.16$\\
    & 0.4  & $-2.44 \pm 0.16$ & $41.16 \pm 0.12$ & $-1.14 \pm 0.14$\\
    & 0.5  & $-2.23 \pm 0.11$ & $41.24 \pm 0.08$ & $-1.23 \pm 0.13$\\
\hline
[\OII]~        & $z$ & $\phi$ & $L_\star$ & $\alpha$ \\
L07 & 0.89 & $-2.25 \pm 0.13$ & $41.33 \pm 0.09$ & $-1.27 \pm 0.14$\\
    & 0.91 & $-1.97 \pm 0.09$ & $41.40 \pm 0.07$ & $-1.20 \pm 0.10$\\
    & 1.18 & $-2.20 \pm 0.10$ & $41.74 \pm 0.07$ & $-1.15 \pm 0.11$\\
    & 1.47 & $-1.97 \pm 0.06$ & $41.60 \pm 0.05$ & $-0.78 \pm 0.13$\\
D13 & 0.35 & $-2.31 \pm 0.24$ & $40.90 \pm 0.18$ & $-1.06 \pm 0.36$\\
    & 0.53 & $-2.85 \pm 0.35$ & $41.13 \pm 0.20$ & $-1.68 \pm 0.36$\\
    & 1.19 & $-2.41 \pm 0.08$ & $41.61 \pm 0.07$ & $-0.95 \pm 0.14$\\
    & 1.46 & $-2.03 \pm 0.05$ & $41.76 \pm 0.05$ & $-0.91 \pm 0.11$\\
    & 1.64 & $-1.68 \pm 0.47$ & $41.73 \pm 0.11$ & $-0.91 \pm 0.11$\\
\hline
[\OIII]~       & $z$ & $\phi$ & $L_\star$ & $\alpha$ \\
L07 & 0.48 & $-2.55 \pm 0.25$ & $41.17 \pm 0.22$ & $-1.49 \pm 0.11$\\
    & 0.42 & $-2.38 \pm 0.22$ & $41.11 \pm 0.24$ & $-1.25 \pm 0.13$\\
    & 0.62 & $-2.58 \pm 0.17$ & $41.51 \pm 0.15$ & $-1.22 \pm 0.13$\\
    & 0.83 & $-2.54 \pm 0.50$ & $41.53 \pm 0.11$ & $-1.44 \pm 0.09$\\
D13 & 0.14 & $-3.67 \pm \infty$ & $41.6 \pm \infty$ & $-1.63 \pm 0.42$\\
    & 0.63 & $-2.57 \pm 0.12$ & $41.44 \pm 0.09$ & $-1.27 \pm 0.11$\\
    & 0.83 & $-2.25 \pm 0.80$ & $41.28 \pm 0.09$ & $-0.76 \pm 0.21$\\
    & 0.99 & $-3.00 \pm 0.23$ & $41.70 \pm 0.13$ & $-0.78 \pm 0.20$\\

\end{tabular}
\caption{Schechter parameters of the observed LF used in this work. Data is from \protect\cite{2007ApJ...657..738L} (L07) and \protect\cite{2013MNRAS.433..796D} (D13); for simplicity, we consider only works that fitted their LF with a Schechter function, neglecting, for example, \protect\cite{2013MNRAS.433.2764G}. We use the raw LF, without dust correction. }
\label{tab:lddata}
\end{table}

\begin{table}
\centering
\begin{tabular}{ccccc}
\hline
H$_\alpha$  & $\phi$ & $L_\star$ & $\alpha$ & $\langle z \rangle $ \\
           & $-2.36 \pm 0.07$ & $41.19 \pm 0.06$ & $-1.33 \pm 0.05$ & $0.4$\\
\hline
\OII~         & $\phi$ & $L_\star$ & $\alpha$ & $\langle z \rangle $ \\
           & $-2.10 \pm 0.03$ & $41.61 \pm 0.02$ & $-1.05 \pm 0.04$ & $1.19$ \\
\hline
\OIII~        & $\phi$ & $L_\star$ & $\alpha$ & $\langle z \rangle $ \\
           & $-2.58 \pm 0.07$ & $41.43 \pm 0.05$ & $-1.30 \pm 0.05$ & $0.72$\\
\end{tabular}
\caption{Mean Schechter parameters of the interopers LF from Tab. \ref{tab:lddata}.}
\label{tab:resfit}
\end{table}

The abundance matching technique required to compute the power-spectrum of interloping lines involves the knowledge of the line LF, which is not as easy as the continuum LF to measure. Fortunately our \Lya signal is only contaminated by interlopers at low redshift ($z < 2$), where observations are more easily available. We use the Schechter LF parameterization \citep{1976ApJ...203..297S} in \cite{2007ApJ...657..738L} and \cite{2013MNRAS.433..796D} (see Tab. \ref{tab:lddata}). The intrinsic intensity of interlopers is not relevant in our work, therefore we use the unprocessed LF, i.e. without dust correction.

Currently the observed interloper LFs are not complete enough to derive a redshift evolution. This forces us to use the variance-weighted mean Schechter parameters to construct the $L=L(M)$ relations at the variance-weighted mean redshift. The same relation is then applied to all redshifts (see CF16). The mean Schechter parameters and redshifts are listed in Tab. \ref{tab:resfit}. In this scenario the redshift evolution of the LF is purely attributed to the halo mass function evolution. Although this might seem a strong assumption, the redshift intervals\footnote{The emission redshift is $1 + z_\mathrm{em} = 	(1+z)(\nu_\mathrm{int}/\nu_{\alpha})$, therefore at $z_{\alpha} = 6,7,8$ the corresponding emission redshifts for the interlopers are $z_{H_\alpha} = 0.30,0.48,0.67$, $z_{\rm OII} = 1.28,1.61,1.94$ and $z_{\rm OIII}= 0.70,0.94,1.19$.}  of the interloper lines that we need to consider are relatively small. For example, for the H$_\alpha$ line (the strongest contributor) the relevant interval is $0.30<z<0.67$. As a result, we believe that the assumption does not affect our conclusions.

\begin{figure*}
\vspace{+0\baselineskip}
{
\includegraphics[width=0.45\textwidth]{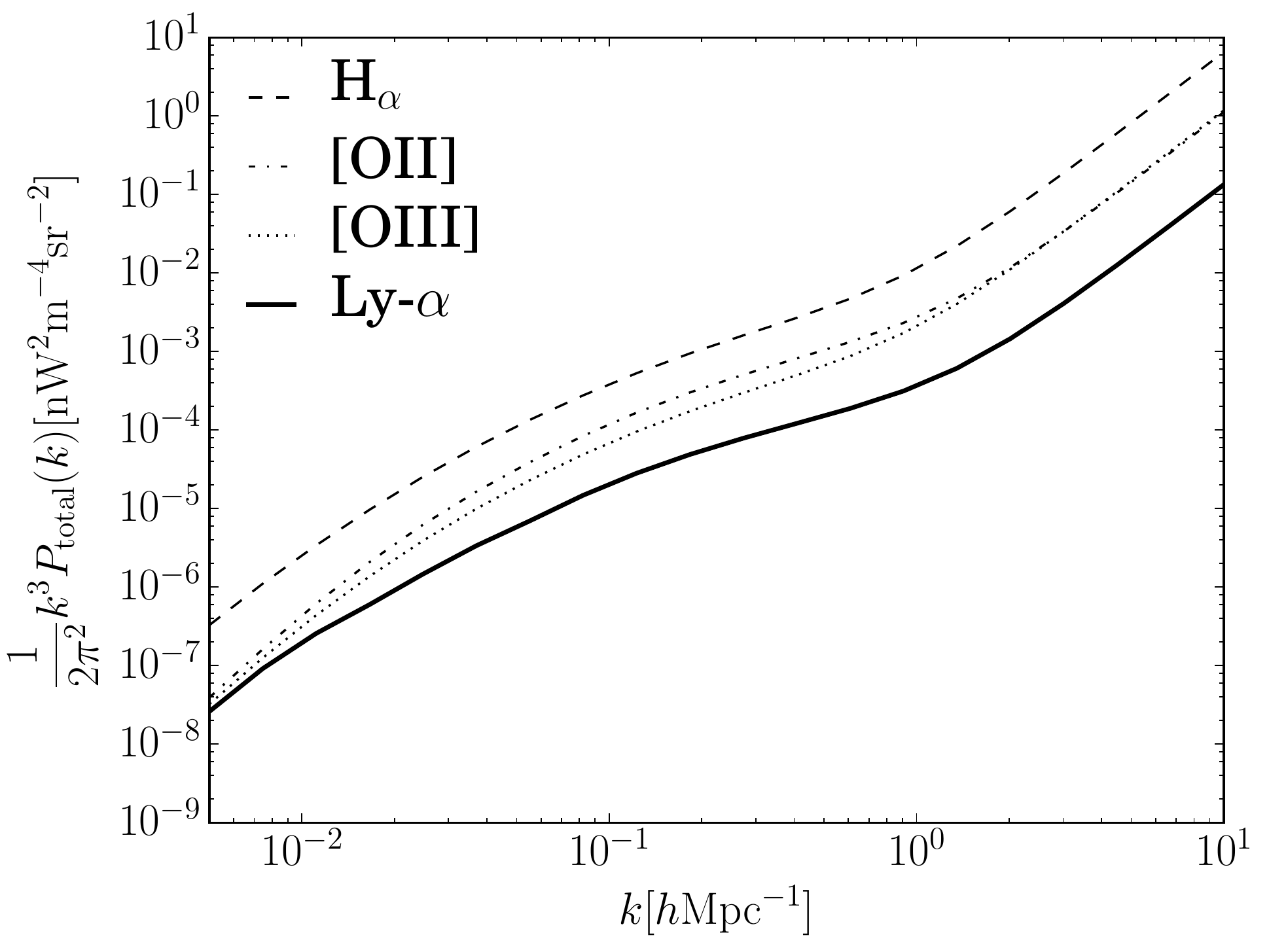}
\includegraphics[width=0.45\textwidth]{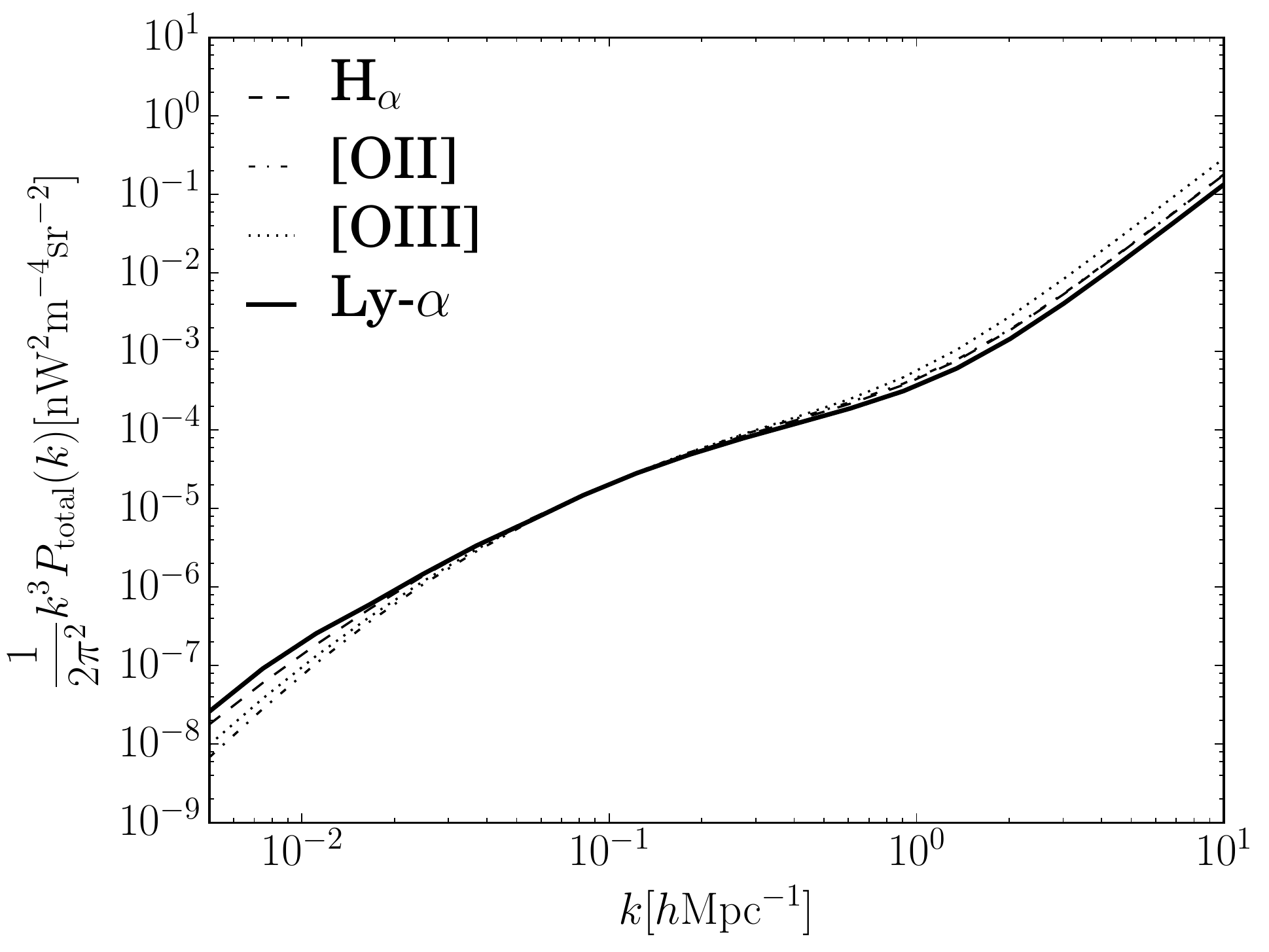}
}
\caption{{\bf Left}: Comparison of the interlopers and \Lya power spectra at $z=7$. {\bf Right}: Same after bright interlopers removal.
}
\label{fig:foreground}
\end{figure*}

Fig. \ref{fig:foreground} shows the PS of interlopers compared with \Lya from $z=7$. As we discuss in Sec. \ref{sec:recap}, incorrectly projecting the interlopers to higher redshift introduces distortions that can amplify their PS. Since the projected interlopers PS is anisotropic, we average it over the solid angle, $P(k) = \frac{1}{4\pi} \int d\Omega P({\bf k})$. However, the anisotropy information can be used to assess the quality of the removal procedure \citep{2014ApJ...785...72G}.  We find that interlopers dominate the PS by 1-2 orders of magnitude on all scales, and that H$_\alpha$ is the dominant confusion source. Therefore an appropriate removal of the interloping PS from \Lya signal, discussed in the following, is crucial. 

\subsection{Interlopers removal}\label{ssec:interrem}
Removing the interloping lines requires a strategy that is different from that used to deal with continuum foregrounds. A possible strategy is to mask the contaminated pixels \citep{2014ApJ...785...72G, 2014ApJ...786..111P, 2015MNRAS.452.3408B}. This is feasible because the galaxy population emitting the interloping lines is very different from the signal sources at EoR: bright galaxies are very rare at high redshift because they are exponentially suppressed in the LF. Hence, if we remove the most luminous pixels from the survey, most of them would be occupied by low-$z$ galaxies and the intensity of interloping lines could be reduced significantly.

However, although straightforward this approach has two drawbacks: (i) if the S/N of the observation is not high, bright voxels can result from noise or foreground fluctuations; (ii) it removes also Ly$\alpha$ flux \citep{2015MNRAS.452.3408B}. For this reason in this work we will use a different approach relying on ancillary galaxy surveys for the identification of the interlopers \citep{2014ApJ...786..111P,  2015ApJ...806..209S,   mappingCII}. This strategy would affect only weakly the Ly-$\alpha$ PS;  however, ancillary surveys have to be sufficiently deep, wide and galaxy redshifts have to be estimated precisely.

To demonstrate the feasibility of such approach, we first perform a calculation similar to that shown in the left panel of Fig. \ref{fig:foreground} but imposing an upper limit to the mass of the interloping galaxies. We assume that the pixels containing galaxies larger than this upper limit are removed from the survey. Fig. \ref{fig:foreground} (right) shows the PS of \Lya signal at $z = 7$ and interlopers, normalizing all power spectra at $k = 0.1~h{\rm Mpc}^{-1}$. Both the mean intensity, the mean bias and the shot noise depend on the upper limit ($2 \times 10^{11}~M_\odot$ for H$_\alpha$, $4.4 \times 10^{11}~M_\odot$ for [\OII],  and $2 \times 10^{12} ~M_\odot$ for [\OIII]).  Removing massive galaxies suppresses very efficiently the PS of the interlopers. We find that the removed voxels occupy only $2$\% of the survey volume.

\begin{figure*}
\vspace{+0\baselineskip}
{
\includegraphics[width=0.45\textwidth]{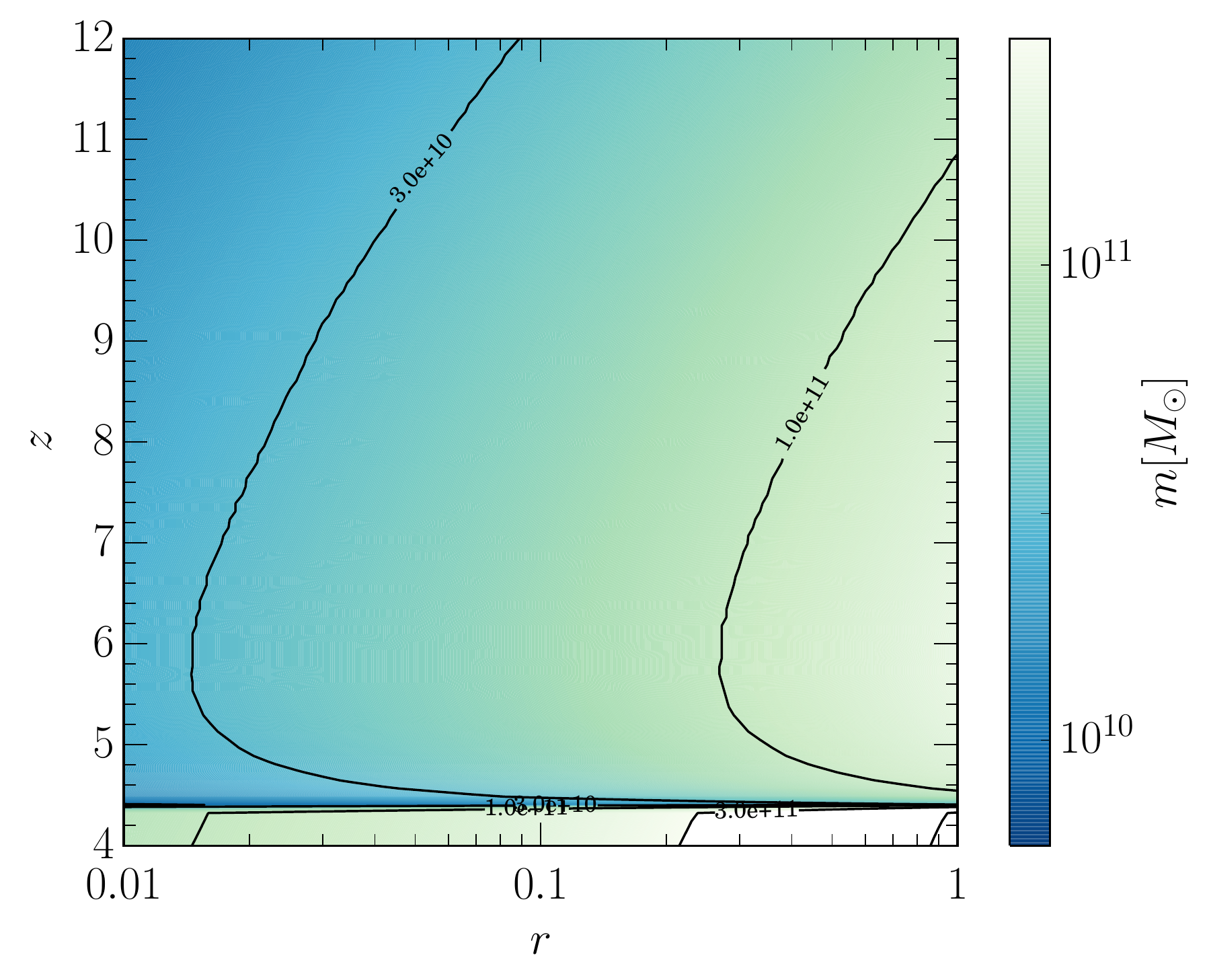}
\includegraphics[width=0.45\textwidth]{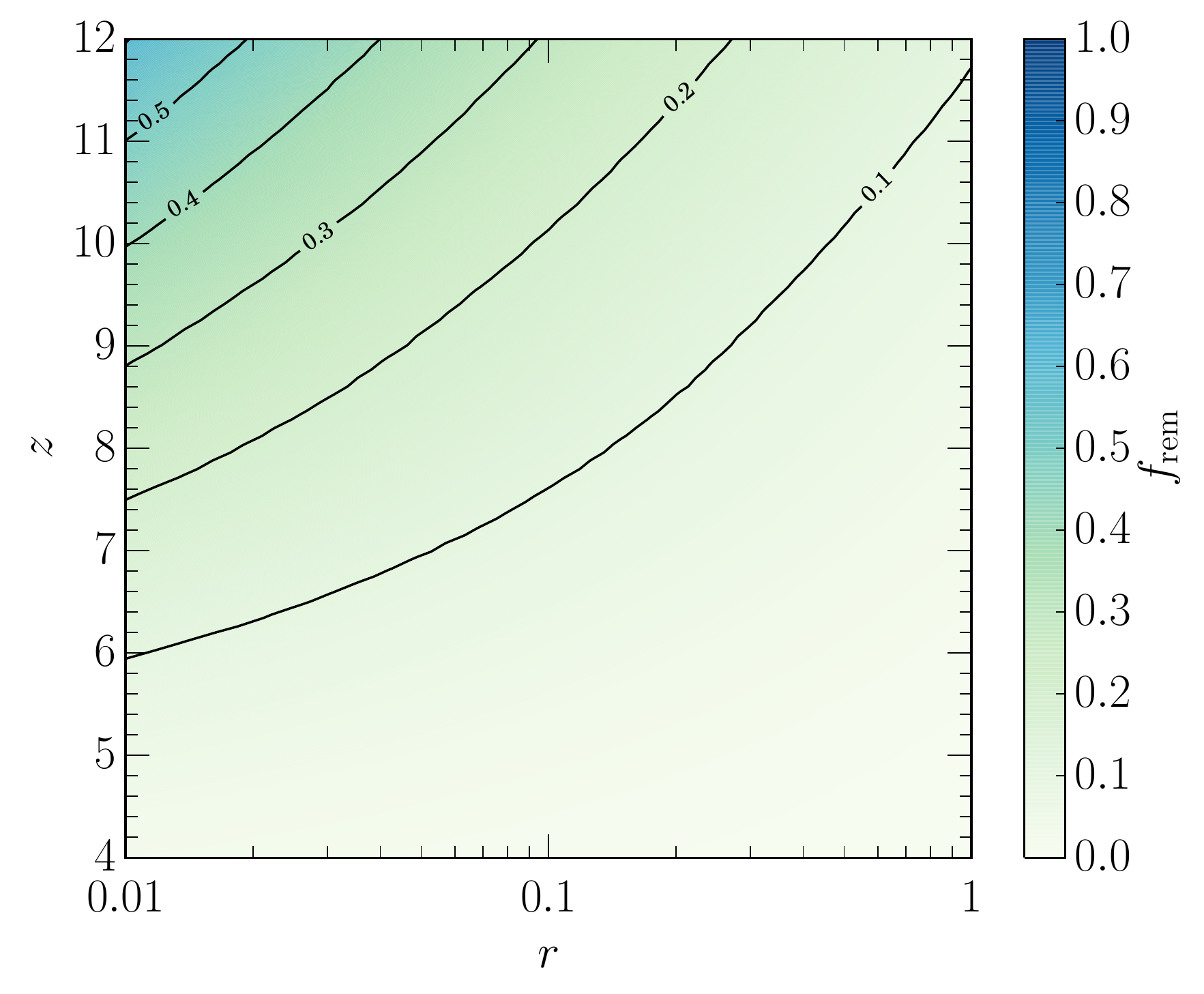}
\caption{\label{fig:frem}
\textbf{Left}: Maximum mass of the galaxies contributing to the interlopers PS. The sharp discontinuity at $z = 4.4$ is due to the H$_\alpha$ line entering the survey. \textbf{Right}: Fraction of voxels that has to be removed to obtain a ratio $r$ between the interlopers and \Lya PS on scale $k = 0.1~h{\rm Mpc}^{-1}$ at redshift  $z$. A redshift uncertainty in the ancillary galaxy survey of 5\% ($R = 20$) has been assumed, thus multiple voxels are discarded for each interloper. 
}
}
\vspace{-1\baselineskip}
\end{figure*}

In the left panel of Fig. \ref{fig:frem} we show the minimum mass of halos that have to be removed from the survey to reach a interloper-to-signal ratio $r$ (defined as the PS ratio at scale $k = 0.1~h{\rm Mpc}^{-1}$) for PS of \Lya from redshift $z$. We find that an effective interloper removal requires to resolve galaxies hosted by halos with $M \simgt 10^{11}M_\odot$ and line flux $f \simgt 10^{-16} {\rm erg~} {\rm cm}^{-2}$. This can be challenging for a large area survey. The fraction of the volume loss can be substantial, as shown by the right panel of Fig. \ref{fig:frem} when considering a 5\% ($R = 20$) redshift uncertainty in the ancillary galaxy survey, resulting in more than one voxels discarded per galaxy. We remind that if more than $\sim30\%$ of the survey volume is masked, the PS reconstruction can be unfeasible \citep{2005Natur.438...45K}. From the right panel of Fig.  \ref{fig:frem} we conclude that cleaning a \Lya IM survey can be intrinsically difficult at $z > 12$, while the volume loss is not problematic for observations at later epochs.  

\begin{figure*}
\vspace{+0\baselineskip}
{
\includegraphics[width=0.45\textwidth]{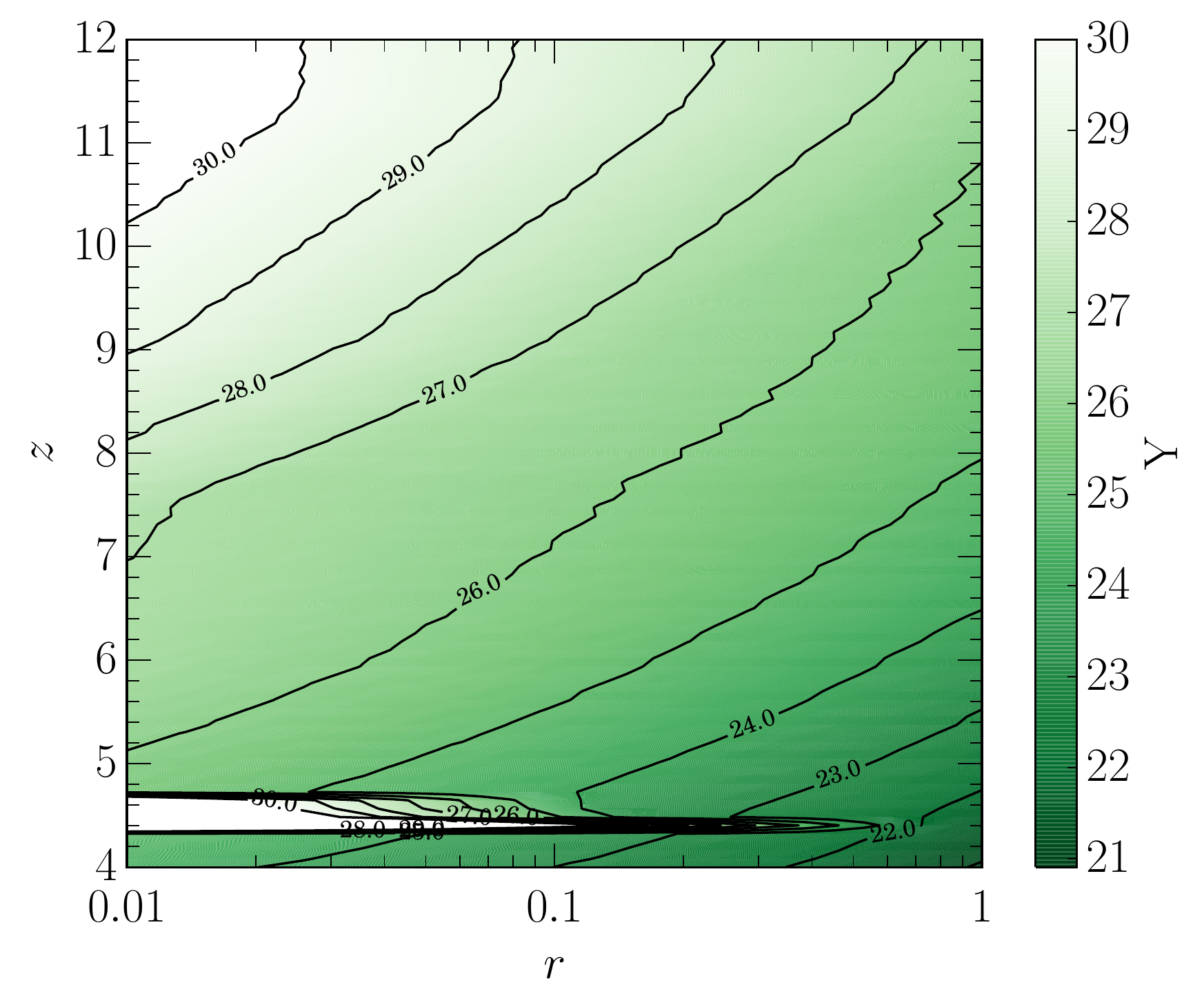}
\includegraphics[width=0.45\textwidth]{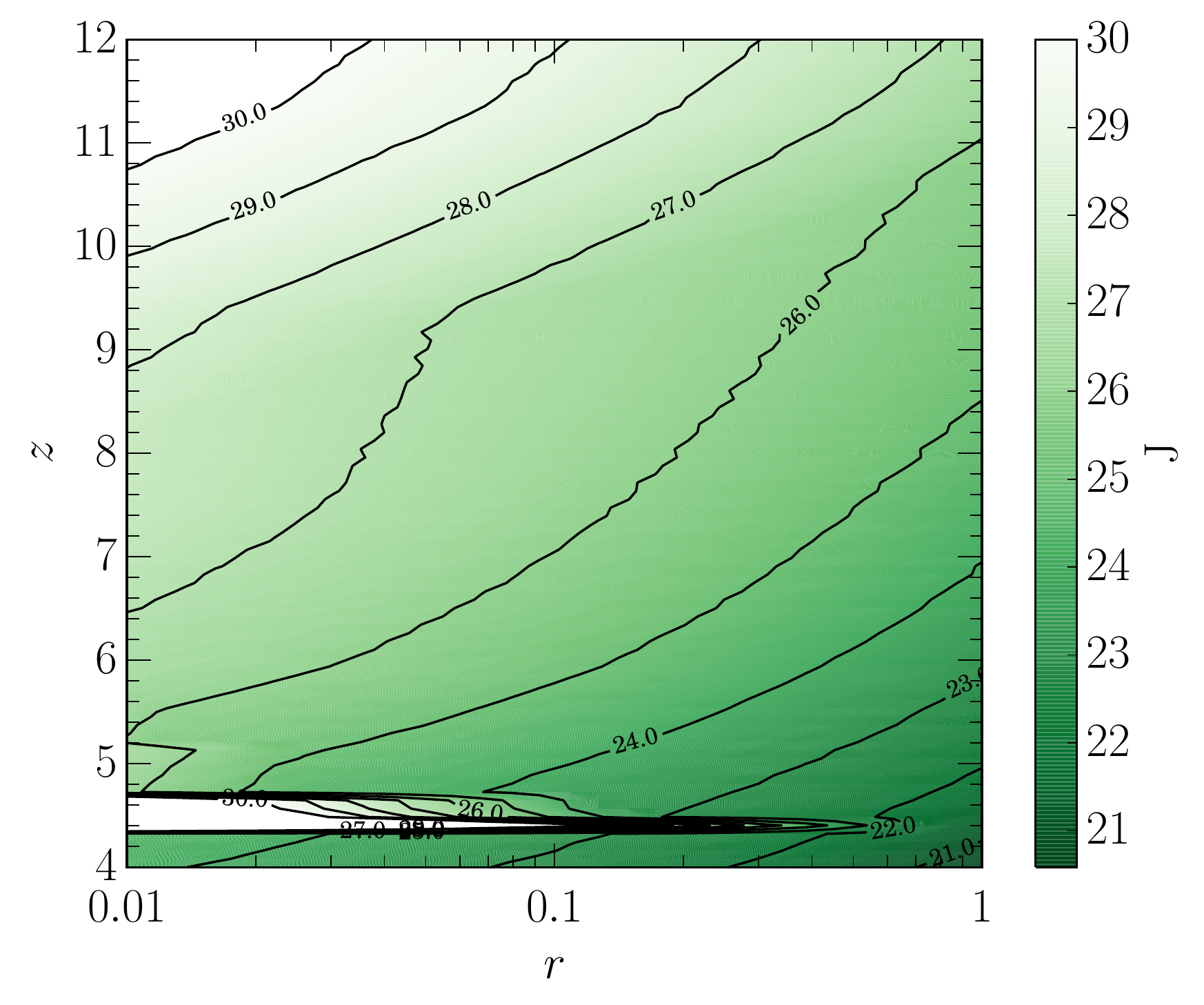}
\includegraphics[width=0.45\textwidth]{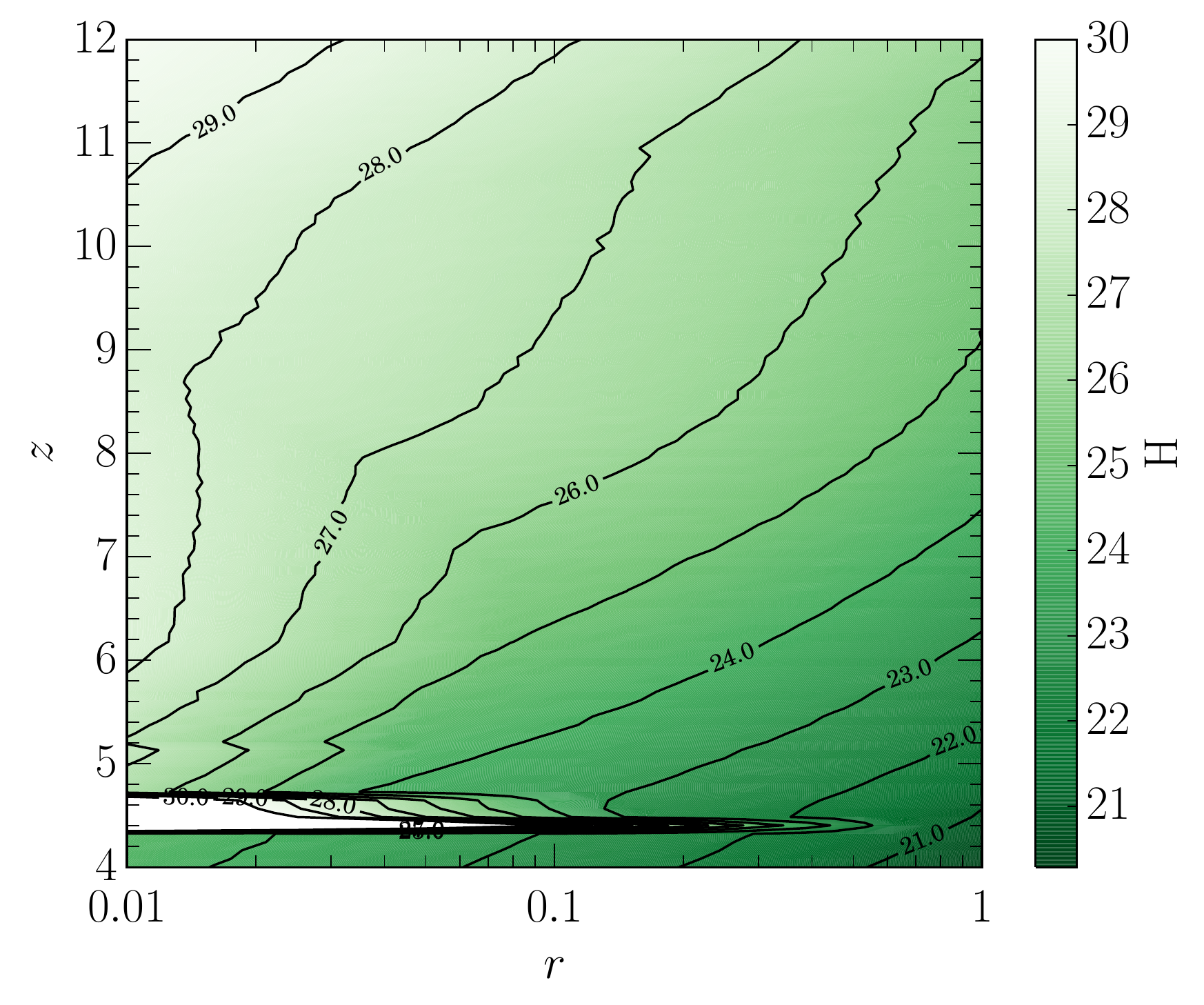}
\includegraphics[width=0.45\textwidth]{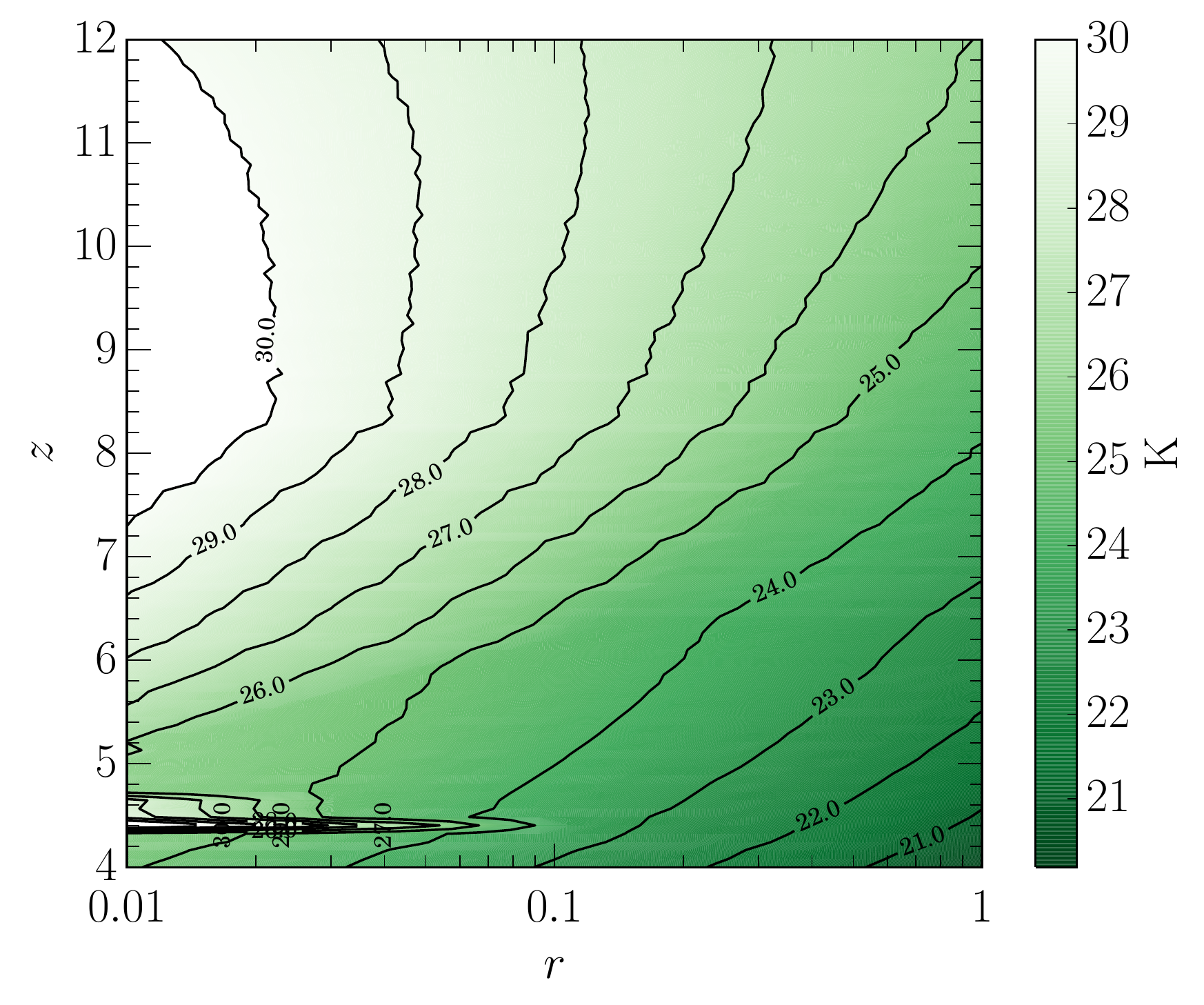}
}
\caption{Limiting magnitude in different spectral bands (Y, J, H, K) required for a survey to clean the \Lya PS signal from interlopers; $r$ is the ratio between the interlopers and \Lya PS on scale $k = 0.1~h{\rm Mpc}^{-1}$ at redshift  $z$. The sharp horizontal feature at $z = 4.4$ is due to the H$_\alpha$ PS from $z = 0$; it  formally diverges (see Eq. \eqref{interamp}).
}
\vspace{-1\baselineskip}
\label{fig:bands}
\end{figure*}

We can then translate the above constraints on a limiting apparent magnitude at which interloper galaxies must be removed. 
To this aim we use the optical and NIR rest frame LFs in \cite{2012ApJ...752..113H} and assign luminosities to DM halos using the abundance matching technique. The apparent AB magnitude at a specified wavelength is obtained from linear interpolation between two neighboring bands in \cite{2012ApJ...752..113H}.
Fig. \ref{fig:bands} shows the maximum depth needed by a survey to remove interlopers as a function of $r$ and signa redshift in the Y, J, H and K bands. To access the signal from late EoR the ancillary survey must reach an AB mag $\simgt 26$.  Compared with the designed sensitivity of future photometric surveys this is rather challenging. For example the EUCLID\footnote{\url{http://sci.esa.int/euclid/}} wide survey will reach a limiting magnitude of $24$ in bands $Y$, $J$ and $H$: this can be enough only to clean the \Lya PS at $z < 4.4$ (without the H$_\alpha$ line). Observing the EoR signal and reaching AB mag $m = 27-28$ is extremely challenging and is at the edge of the capabilities of future instruments, such as WFIRST\footnote{\url{http://wfirst.gsfc.nasa.gov}} or FLARE. 

\section{Cross-power spectra}
\label{sec:cps}
 
In Sec. \ref{ssec:interrem} we have discussed an interloper removal method based on ancillary surveys. In spite of the optimistic assumptions (for example, we have neglected the scatter in the line luminosity, SFR and halo mass relations) the required masking depth is relatively demanding. 

An alternative strategy would be to use the cross-correlation between two different intensity mapping experiments contaminated by different interloping lines. The $157.7~\mu$m~[\CII]~ fine structure line is the brightest of all the metal lines, contributing generally up to $\sim1\%$ of the total galaxy IR luminosity. Its line luminosity  scales tightly with the SFR, but is affected also by the ISM metallicity \citep{2013MNRAS.433.1567V,2015ApJ...813...36V}. The removal of continuum foreground and interloping lines for [\CII]~auto-correlation PS measurements was investigated in \citet{mappingCII}. In this section we investigate its cross-correlation with the \Lya line. The interloping lines for these two signals are not correlated with each other because they are produced in non-overlapping redshift intervals.

\subsection{[\CII] line intensity and cross power}

The mean [\CII]~intensity can be directly obtained from the galaxy line luminosity \citep{ 2016MNRAS.455..725C}:
\begin{equation}
I_{\rm CII}(z)=\frac{c}{4\pi \nu_{\rm CII} H(z)} \int d{M} \frac{d{n}}{d{M}} L_{\rm CII}(M, z).
\end{equation}
It spatially fluctuates following the large scale DM density field multiplied by a line luminosity-weighted mean bias, $\langle b \rangle_{\rm CII}$:
\begin{equation}
\delta I_{\rm CII} = I_{\rm CII} \langle b \rangle_{\rm CII} \delta;
\end{equation}
where $\delta$ is the DM density contrast, 
\begin{equation}
\langle b \rangle_{\rm CII} =   \frac{1} { \rho_{\rm CII} } \int d{M} \frac{d{n}}{d{M}} b(M, z) L_{\rm CII}(M, z).
\end{equation}

The cross-correlation PS includes three main terms:
\begin{itemize}
\item {\em Large scale DM fluctuations} originating from the \Lya and [\CII]~ lines, both emitted by the ISM. This component  dominates the PS on scales $\simgt 1$~Mpc. It can be written as
\begin{equation}
P_{{\rm CII}, \alpha}^{\rm s,s}(k, z) = I_{\rm CII}(z) I_{\alpha}^{\rm s}(z) \langle b \rangle_{\alpha} \langle b \rangle_{\rm CII} P_{\rm dm}(k,z),
\end{equation}
where $I_\alpha^{\rm s}$ is the \Lya emission from the ISM;
\item {\em Fluctuations from UV continuum emission} resulting from the correlation between \Lya emission in the IGM and [\CII]~ emission in the ISM. \Lya fluctuations are produced by (i) UV emission from the galaxies, and (ii) Lyman absorption followed by relaxation in the IGM.  We can express the spatial intensity fluctuations as 
\begin{multline}
\label{Icont}
\delta I^\alpha_{\mathrm{cont}}(z) = \frac{c h_P \nu_{\alpha}}{4 \pi (1+z)} \sum_{n = 2}^{\infty} P_{abs}(n, z) f(n)  \times \\
\times \int dz' \frac{\dot n_\nu(\nu', z')} {H(z')} \prod_{n' = n+1}^{n'_{max}} T(n', z_{n'}) [\langle b(z')\rangle_{\nu'} \delta]= \\
=\frac{1}{4\pi} \int d\Omega \int_z^{+\infty}dz' A(z, z') \dot n_{\nu}(\nu', z')[\langle b(z')\rangle_{\nu'} \delta],
\end{multline}
where $P_{abs}(n, z)$ is the IGM absorption probability of a Lyman-$n$ photon at redshift $z$, $f(n)$ is the fraction of \Lya photons emitted by an HI atom during the decay from the $n$-th energy level, $\dot n_\nu$ is the number of UV photons emitted per unit time, volume and frequency, $T(n, z) = 1 - P_{abs}(n,z)$ is the transmission probability; we refer to CF16 for details. The associated cross-correlation PS is
\begin{multline}
P_{{\rm CII}, \alpha}^{s,c}(k, z) =\\
\left[I_{\rm CII}(z) \langle b \rangle_{\rm CII} \int_z^{+\infty} d{z'} A(z, z') \dot n_\nu'D'\frac{\sin(kl')}{kl'}\right] P_{\rm dm}(k,z),
\end{multline}
where $l'(z, z') = c \int_z^{z'} dx~ H(x)^{-1}$. It becomes important only on scales $\gsim 100$~Mpc as fluctuations on scales smaller than the typical mean free path of a photon with energy between the \Lya and the Lyman-limit are washed out.
\item {\em Shot noise} due to the discrete nature of the sources dominates on small scales:
\begin{equation}
P_{{\rm CII}, \alpha}^{\rm SN}(k, z) = I_{\rm CII} I_{\alpha}^{\rm h} \frac{1}{  \rho_\alpha^{\rm h} \rho_{\rm CII}  }      \int dM \frac{d{n}}{d{M}} L_\alpha(M) L_{\rm CII}(M);
\end{equation}

\end{itemize}

\begin{figure}
\vspace{+0\baselineskip}
{
\includegraphics[width=0.45\textwidth]{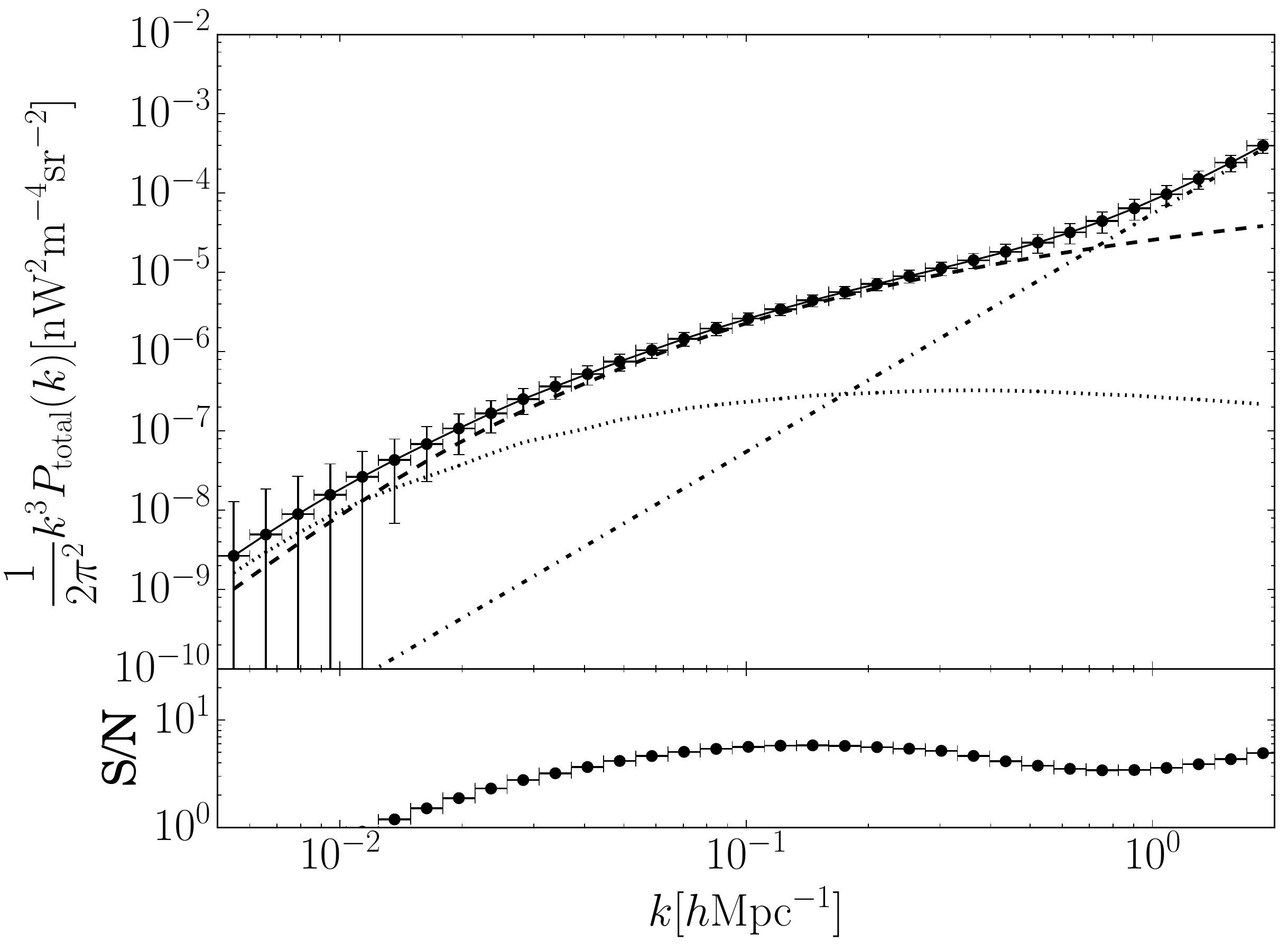}
}
\caption{\textbf{Top}: Ly$\alpha$-[\CII] cross-correlation power spectrum at $z = 6$. We show the total PS (solid), large scale dark matter fluctuations (dashed), fluctuations from UV continuum emission (dot-dashed),  and shot noise (see text). The error bars are computed considering the same [\CII] instrumental setup in \protect\cite{mappingCII} and in Sec. \ref{ssec:foreobs}. \textbf{Bottom}:  S/N of the observation as a function of wavenumber.}
\label{fig:lyacii7}
\end{figure}

Fig. \ref{fig:lyacii7} shows the Ly$\alpha$-[\CII] cross-correlation PS at $z=6$ with the three main components plotted separately. As expected the PS is largely dominated by the ISM emission, and only on scales  $\gsim100$~Mpc the IGM becomes important. We plot also the S/N of an hypothetical observation, using the same [\CII] survey proposed in \cite{mappingCII}. For consistency, we adopt a spectral resolution $R=100$ and angular resolution $\Delta \theta = 42$ arcsec for both [\CII]  and \Lya observations. The total survey area is $250~{\rm deg}^2$, corresponding to about $100$ pointings, each with exposure time of $10^5$~s (total observing time $10^7$s, or about 4 months).   

\begin{figure*}
\vspace{+0\baselineskip}
{
\includegraphics[width=0.45\textwidth]{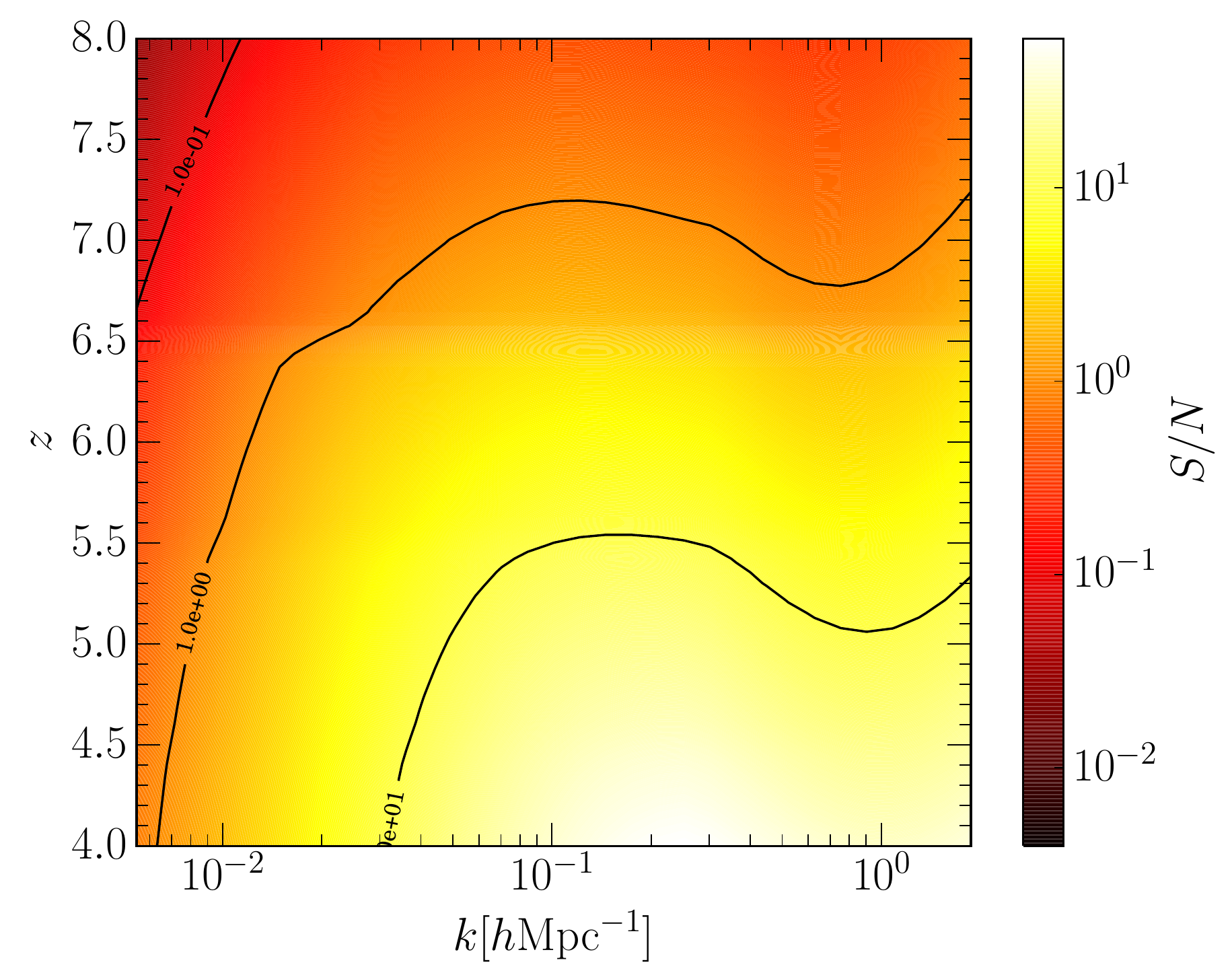}
\includegraphics[width=0.45\textwidth]{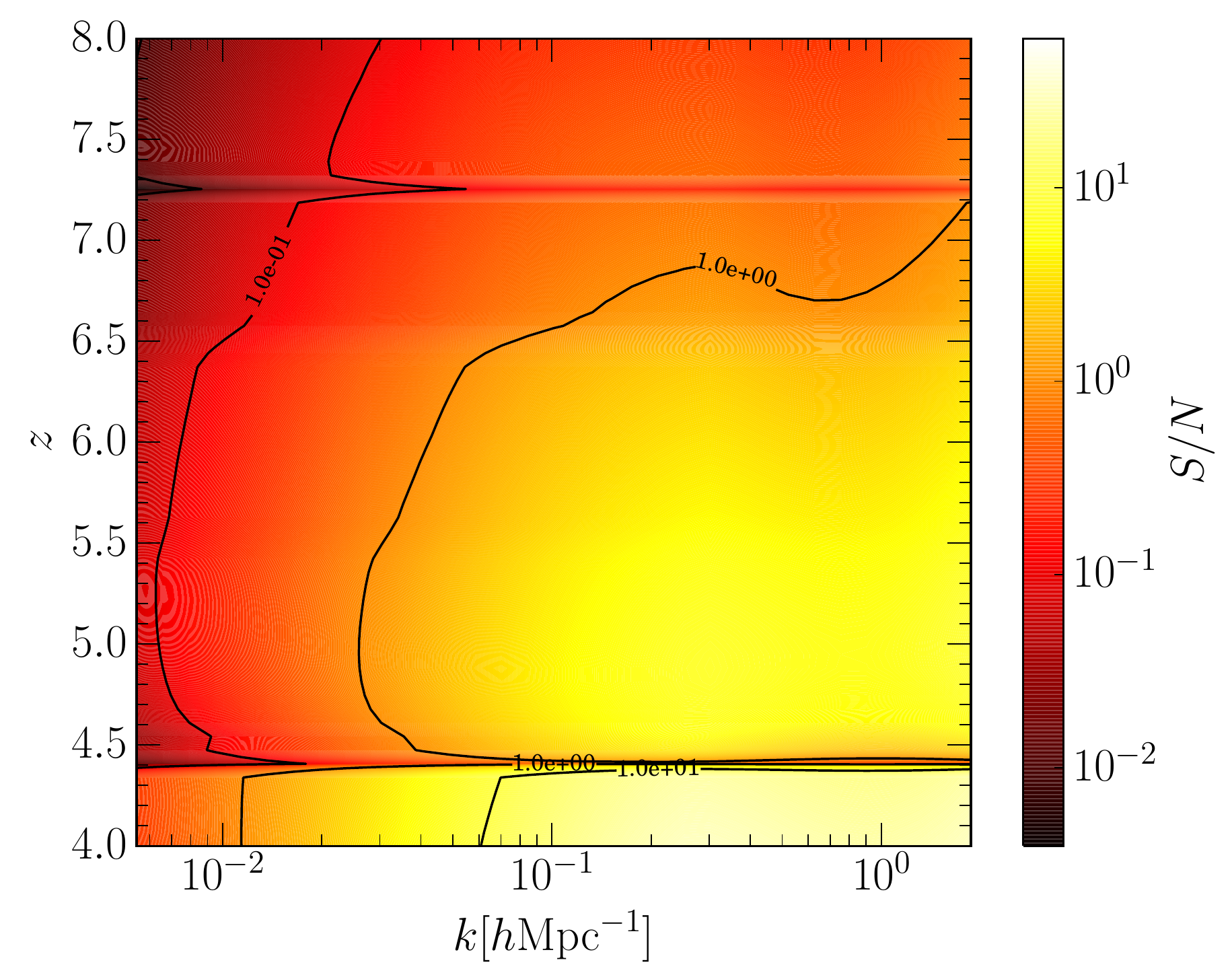}
}
\caption{\textbf{Left:} S/N of the Ly$\alpha$-[\CII] cross-correlation spectrum vs. wavenumber after interloping lines have been removed; \textbf{Right:} Same as left panel, before interloper removal. The observational setup is the same as in  Fig. \ref{fig:lyacii7}}.
\vspace{-1\baselineskip}
\label{fig:ciisn}
\end{figure*}

The left panel of Fig. \ref{fig:ciisn} shows the S/N  as a function of $k$ and $z$ (assuming $\Delta k = 1.2 k$) for the most optimistic case where all interlopers are cleanly removed (Eq. \eqref{crvar} with $P^i_{f,1} = P^i_{f,2} = 0$). 
These results are encouraging, because they show that in principle a Ly$\alpha$-[\CII] intensity mapping observation of the late EoR is feasible.

However, as we showed in Sec. \ref{ssec:foreobs}, interlopers can increase the PS variance well beyond the instrumental noise. 
For [\CII]~IM, CO rotational lines (see \citealt{2014ApJ...794..142G,2009MNRAS.399..264B,2014MNRAS.444.1301P} for detailed CO emission line studies) are the most important interlopers. They have PS amplitude comparable or even larger than the [\CII] one, and therefore they must be removed (see \citealt{mappingCII,2015ApJ...806..209S}). 

We then added the CO lines, H$_\alpha$, [\OIII] and [\OII] lines to the variance of the cross-correlation PS (see Eq. \ref{crvar}). 
The right panel of Fig. \ref{fig:ciisn} shows the S/N with interlopers (the strong features at $z \approx 4.5$ and $7.2$ are due to H$_\alpha$, CO 2-1 and CO 3-2 lines entering the survey, respectively).
The effect of the interlopers is to decrease the S/N significantly; without an efficient removal the EoR signal is inaccessible.  However, compared to the \Lya auto-correlation PS discussed in Sec. \ref{ssec:interrem}, the  Ly$\alpha$-[\CII] cross-correlation spectrum can be more easily recovered by using  a shallower ancillary survey within the capability of a near future instrument. To support this statement we recompute the S/N however removing all the interlopers with $m_{\rm AB} < 24$ (vs. $\sim26$ for recovering the \Lya auto-correlation PS) in the EUCLID NIR bands (Y, J and H), finding that the recovered signal matches almost perfectly the model without interlopers. 

This approach, even though promising, is more difficult to interpret. The information recovered by the cross-PS is degenerate and it is not possible to recover information about \Lya or \CII lines individually. It is necessary to rely on ancillary data to extract the relevant astrophysical information, such as a PS measurement of one of the two lines or a combination of several cross-correlations. Another possibility is to cross-correlate with resorved sources, such as QSOs \citep{2015arXiv150404088C} or LAE (Comaschi \& Ferrara in prep.).  

\section{Conclusions}
\label{sec:conc}
We have investigated the feasibility of a \Lya intensity mapping experiment targeting the collective signal from galaxies located at $z > 4$. We have used a recently developed analytical model to predict the \Lya power spectrum, and carefully studied the main observational challenges. These are ultimately quantified by the expected S/N for various observational strategies.

We found that in principle the \Lya PS for $z<8$ is well at reach of a small space telescope (40 cm in diameter); detections with low S/N are possible only in some optimistic cases up to $z\sim10$. However, the foreground from interloping lines represent a serious source of confusion and must be removed. The host galaxies of these interloping lines can be resolved via an ancillary photometric galaxy survey in the NIR bands (Y, J, H, K). If the hosts are removed down to AB mag $\sim26$, then the \Lya PS for $5 < z < 9$ can be recovered with good S/N. We further found that, by cross-correlating the \Lya emission with [\CII]~emission from the same redshift,  the required  depth of the ancillary galaxy survey could be is within reach of Euclid (AB mag $\sim24$). 

The results of this work show the yet unexplored, remarkable potential of \Lya IM experiments. By using a small space telescope and a few days observing time it is possible to probe galaxies hosted by DM halos with $M \approx 10^{10}~M_\odot$ well into the EoR. Such galaxies emit the bulk of the collective \Lya radiation. However, the technical difficulty is represented by the interloping lines removal, which sets demanding requirements to the ancillary survey: the combination of very large survey areas ($\sim250$ deg$^2$) and significant depth (AB mag $\sim26$) appear to be challenging also for the next generation telescopes. We have suggested however, that such problem can be overcome by cross-correlating the \Lya IM with other lines (as the $157.7~\mu$m~[\CII]~ fine structure line), thus making a strong synergy between programs targeting different bands almost mandatory.


\bibliography{paper2}

\appendix

\section{Power Spectrum variance}
\label{app:fullcalc}

In this Appendix we discuss the derivation of the deviate of Eq. (\ref{varianceline}). For simplicity we consider only two components: the line intensity $I_\alpha$ and the detector noise $I_{\rm N}$; we will work in $\bf k$-space:  
\begin{equation}
\delta I = \delta I_\alpha + \delta I_{\rm N}.
\end{equation}
Since other components do not correlate with $I_\alpha$ and with $I_{\rm N}$, adding them to the results is trivial.

In this paper we use the Fourier convention from \cite{1993sfu..book.....P}:
\begin{gather}
\delta_{\bf k} = \int_V \delta({\bf x}) e^{-i {\bf k}\cdot {\bf x}} d^3 {\bf x};
\end{gather}
$\delta_{\bf k}$ has a Gaussian Probability Distribution function (PDF). If $\delta_{\bf k} $ is written in  polar coordinates, $\delta_{\bf k} = r_{\bf k} \exp{i\phi_{\bf k}}$, the PDF assumes the form
\begin{equation}
\label{pdf}
g_{\bf k}(r_{\bf k}, \phi_{\bf k}) d{r_{\bf k}} d{\phi_{\bf k}} = \frac{2 r_{\bf k} d{r_{\bf k}}}{\sigma_{\bf k}^2} \left( \frac{d{\phi_{\bf k}}}{2\pi} \right)e^{-\frac{r_{\bf k}^2}{\sigma_{\bf k}^2}}.
\end{equation}
With Eq. \eqref{pdf} it is easy to prove that $\langle \delta_{\bf k}\delta^*_{\bf p} \rangle = \eta_{\bf kp}\sigma_{\bf k}^2$ (where $\eta_{\bf kp}$ is the Kronecker delta function).  The cases for  $\delta I_\alpha$ and for $\delta I_N$ are similar with the only exception that the variance of $\sigma_N$ does not depend on ${\bf k}$.

Expanding the first term in Eq. \eqref{ptotvar}, we get
\begin{align}
\langle ( \delta I \delta I^*)^2 \rangle &= \langle \mid \delta I_\alpha \mid^4 \rangle + \langle \mid \delta_{\rm N} \mid^4 \rangle + {\langle \delta I_\alpha^2 (\delta I^*_{\rm N})^2 \rangle} + \nonumber \\ 
&+ { \langle (\delta I^*_\alpha)^2 \delta I_{\rm N}^2 \rangle} + 4\langle \mid \delta I_\alpha \mid^2 \mid \delta I_{\rm N} \mid^2 \rangle + \nonumber \\
&+ {2 \langle \mid \delta I_\alpha \mid^2 \delta I_\alpha \delta I_{\rm N}^* \rangle} + {2\langle \mid \delta I_\alpha \mid^2 \delta I_\alpha^* \delta I_{\rm N} \rangle} \nonumber \\
&+ {2 \langle \delta I_\alpha \mid \delta I_{\rm N} \mid^2 \delta I_{\rm N}^* \rangle} +{2 \langle \delta I_\alpha^* \mid \delta I_{\rm N} \mid^2 \delta I_{\rm N} \rangle } \nonumber \\
&= \langle \mid \delta I_\alpha \mid^4 \rangle + \langle \mid \delta_{\rm N} \mid^4 \rangle +4\langle \mid \delta I_\alpha \mid^2 \mid \delta I_{\rm N} \mid^2 \rangle 
\end{align}
where terms like $\langle (\delta I^*_\alpha)^2 \delta I_{\rm N}^2 \rangle$ are null because of the averaging over the phase $\phi$ in Eq. \eqref{pdf}
\begin{equation}
  \langle (\delta I^*_\alpha)^2 \delta I_{\rm N}^2 \rangle \propto \int d \phi_\alpha d \phi_{\rm N} e^{-2 i \phi_\alpha} e^{2i \phi_{\rm N}} = 0.
\end{equation}

The second term in Eq. \eqref{ptotvar} is 
\begin{equation}
\langle ( \delta I \delta I^*) \rangle^2 = \left( \langle \mid \delta I_\alpha \mid^2 \rangle + \langle \mid \delta_{\rm N} \mid^2 \rangle \right)^2.
\end{equation}

Using the fact that both $\delta I_\alpha$ and $\delta I_{\rm N}$ are Gaussian we have
\begin{equation}
\langle \mid \delta_{\bf k} \mid^4 \rangle = \int r_{\bf k}^4 \frac{d r_{\bf k}^2}{\sigma_k^2} e^{-\frac{r_{\bf k}^2}{\sigma_k^2}} = 2 \sigma_k^4,
\end{equation}
and finally
\begin{equation}
\langle ( \delta I \delta I^*)^2 \rangle - \langle ( \delta I \delta I^*) \rangle^2 = \left( \langle \mid \delta I_\alpha \mid^2 \rangle + \langle \mid \delta_{\rm N} \mid^2 \rangle \right)^2.
\end{equation}

\end{document}